\documentclass[12pt]{article}

\usepackage{graphicx}
\usepackage{subfigure}
\usepackage{amsmath}
\usepackage{amsfonts}
\usepackage{amssymb}
\usepackage{calligra}
\usepackage{bbold}

\def\T{\mathcal{T}}
\def\p{\partial}
\def\O{\mathcal{O}}

\def\d{\partial}

\def\mpl{{M_{\rm{Pl}}}}

\def\be{\begin{equation}}
\def\ee{\end{equation}}

\def\d{\partial}

\def\mpl{{M_{\rm{Pl}}}}

\def\lsim{\mathrel{\rlap{\lower3pt\hbox{\hskip0pt$\sim$}}
     \raise1pt\hbox{$<$}}}         
\def\gsim{\mathrel{\rlap{\lower3pt\hbox{\hskip0pt$\sim$}}
     \raise1pt\hbox{$>$}}}         

\newcommand{\comment}[1]{}
\newcommand{\Expect}[1]{\left\langle #1 \right\rangle}

\begin{document}

\numberwithin{equation}{section}

\begin{flushright}

{NYU-TH-10/27/2015}
\end{flushright}

\vskip 1cm

\begin{center}

{\Large {Metamorphosis of the Cosmological Constant \\ \vskip 0.2cm

and 5D Origin of the Fiducial Metric}}

\end{center}

\thispagestyle{empty}

\vskip 0.5 cm

\begin{center}

{\large  {Gregory Gabadadze and Siqing Yu}}

 \vspace{.2cm}

{\it Center for Cosmology and Particle Physics,
Department of Physics,}
\centerline{\it New York University, New York,
NY, 10003, USA}

\end{center}

\vskip 1cm

\begin{abstract}

In a recently proposed theory, the cosmological constant (CC)  does not curve spacetime in our universe, 
but instead gets absorbed  into another universe endowed with its own dynamical metric, nonlocally 
coupled to ours. Thus, one achieves a long standing  goal of removing   entirely  any cosmological 
constant from our universe.  Dark energy then cannot be due to a 
cosmological constant, but must be obtained  via other mechanisms. Here we focus on 
the  scenario in which dark energy is due to  massive gravity and  its extensions.  
We show how the metric of the other universe, that absorbs our CC, also gives rise to the 
fiducial metric  known to be necessary for the diffeomorphism invariant 
formulation of massive gravity.  This is achieved in a framework where  the other 
universe   is described by 5D AdS gravity, while our universe lives on its boundary and is endowed
with dynamical massive gravity.  A non-dynamical  pullback of the bulk AdS metric  acts as the fiducial 
metric  for  massive gravity on the boundary.  This framework also removes a  
difficulty caused by the quantum strongly coupled behavior of  massive gravity at the 
$\Lambda_3$ scale: in the present approach, the massive gravity action 
does not receive  any loop-induced counterterms, despite being strongly coupled.  

\end{abstract}

\newpage

\section{An Unconventional Path}\label{intro}

The importance of the cosmological constant (CC) problem  is well-known, 
so are the difficulties in solving it (see \cite{Weinberg:1988cp,Polchinski:2006gy}, and references therein). 
It is clear that unconventional approaches are needed. 
 Along these lines, Tseytlin \cite{Tseytlin:1990hn} had made an interesting  proposal  
building  on an earlier idea of \cite{Linde:1988ws},   and inspirations 
from T-duality of  string theory. He suggested  to apply the  least action principle 
to the ``volume normalized action," ${\bar S}$:
\begin{equation}\label{ac1}
\bar{S}\equiv \frac{S}{V_g}=\frac{\int d^4x\sqrt{|g|}(R+2\mathcal{L}_{\rm SM})}{\int d^4x\sqrt{|g|}},
\end{equation}
instead of that principle being applied to ${S}$.
Here,  $V_g= \int d^4x\sqrt{|g|}$ is the invariant space-time volume in the  
$16\pi G_N=1$ units (it is assumed that  $V_g$ is regularized, as in Section 5).  
The above modified action should be thought 
as a certain low energy effective action, hopefully emerging from more conventional 
high energy physics \cite{Tseytlin:1990hn}.    Since the CC problem is a  low energy 
problem, it is  reasonable  to expect that its  solution will not be influenced by  
an exact form of the high energy completion of (\ref {ac1}).   

Furthermore, $\mathcal{L}_{\rm SM}$ in (\ref {ac1}) denotes  
a Lagrangian of all the fields of nature but gravity, coupled to gravity.   
As argued in \cite{Gabadadze:2014rwa}, for consistency with the empirical data 
the Lagrangian  $\mathcal{L}_{\rm SM}$ needs to be regarded as a quantum effective  Lagrangian  
coupled to classical  gravity.  This classical   gravity will be subsequently quantized using (\ref {ac1}).
Such an unconventional procedure of quantization will be reviewed 
in Section \ref{secloop}, where earlier works that made 
this procedure more precise will be referenced. Till then we will not use 
any specific form of $\mathcal{L}_{\rm SM}$, but discuss only its constant part. 

It is straightforward to see that a cosmological constant is an unphysical 
parameter in (\ref {ac1}):  a shift of $\mathcal{L}_{\rm SM}$ by any constant changes  
$\bar S$ by an additive constant, and the latter does not affect the 
equations of motion obtained  by varying $\bar S$.   

While this appears to be an efficient way to get rid of  an arbitrary CC irrespective of its origin, the equations 
of motion obtained  from $\bar S$ reveal the high cost of the proposed solution: in the 
equations, the local Ricci curvatures are determined by 
spacetime averages over past and future volumes, when these 
volumes are well-defined \cite {Tseytlin:1990hn}. This seems to be  
a dramatic departure away from the conventional local field theory paradigm.

Such spacetime nonlocality (and therefore,  acausality \cite{ArkaniHamed:2002fu}) is however operative only for a
cosmological constant, while  all the local interactions that  can be  measured in a laboratory -- even if 
in one with the size of the observed universe -- are not affected. 
One may wonder how the mechanism could distinguish  between a cosmological constant
and, say,  a very flat scalar potential.  It does so by being nonlocal in time, and hence being   able 
to target and eliminate  only the cosmic fluid that  does not red-shift in future infinity \cite{ArkaniHamed:2002fu}.  
As a result, a CC is eliminated,  while 
the slow roll inflation, radiation,  and matter dominated epochs of cosmology remain intact.\footnote{The mechanism is applicable to a single vacuum universe  \cite{Tseytlin:1990hn,Linde:1988ws}. Its generalization to the case that adopts chaotic inflation with multiple vacua  
is not known to the authors. Throughout the paper we focus on a scenario  in which our 
4D universe evolves in a single vacuum state.} 

What about non-cosmological solutions of Einstein's gravity?  Since the spatial 
nonlocality of the proposed equations  is 
operative only for sources that  have an infinite spatial support, then 
intact remain also all spatially local solutions of Einstein's gravity. Thus, the high cost  
incurred by  the solution of the CC problem, while somewhat bothersome, remains  undemanding,   
at least until inconsistencies of the proposal are found.

We add  that the  action   (\ref {ac1})   can readily be used to obtain the 
classical Hamiltonian of the theory: The multiplier, 
$1/V_g$,   does not contain  any  derivatives of the metric, and 
affects the determination of the  canonical momenta only by rescaling them with 
a multiplicative numerical factor of $1/V_g$.  

To summarize,  in the context of classical gravity and conventionally 
quantized particle physics,  if  $\bar S$ is used as a gravitational action for a single  
vacuum state in which the universe is presumed to evolve,   
then the CC is removed,  without any  apparent contradiction 
with  observations  (see also \cite {Davidson},  for further 
independent work on the proposal of \cite{Tseytlin:1990hn}\footnote{The name 
used  for such theories in  \cite {Davidson}, the Normalized General Relativity (NGR), 
is  better reflecting  the fact  that the   non-locality  of such theories 
is  of a different nature  from the ones ordinarily considered;  
for instance, equations of motion are only  amended by a constant-valued functional.}).

However, the trouble comes with the graviton quantum loops \cite{Tseytlin:1990hn}.
This problem emerges in  the low energy theory,  even before we ask
the question of the existence of a full-fledged UV complete quantum theory of gravity, from which 
the action (\ref {ac1}) might originate. To see this, we recall that the quantization  of  (\ref {ac1})  
would use  the path integral with the kernel 
\be
{\rm exp}\left ( {i {\bar S \over {\hbar}} } \right )\, \equiv\,
\int_{-\infty}^{+\infty} d \lambda\, d\tau \,  {\rm exp} \left \{ {i \over \hbar} \left [ \frac{1}{\tau}S+\lambda(V_g-\tau) 
\right ] \right \},
\label{expS}
\ee
where  we have introduced  integrations w.r.t.  $\tau$ and $\lambda$,
which are real parameters.\footnote{The $\lambda$ integration is needed to determine $\tau$, while 
the $\tau$ integration determines $\lambda$; the latter enters the metric equations of motion.} 
It is clear  from (\ref {expS})  that  the new Planck constant that governs the loop expansion for  gravity  is $\hbar \tau$.
This quantity is infinite (or at least very large in units of the Planck length),  since $\tau =V_g$ is infinite 
(or at least as large and old as the present-day universe); therefore, the loop expansion in $\hbar \tau$  
would diverge and  ruin the aforementioned classical solution of the big CC problem \cite{Tseytlin:1990hn}. 
Even if a resummation of the loop expansion were found, the result would significantly differ  
from (\ref {ac1}), and one would have to work anew for a solution of the CC problem.  There is no reason to expect 
that the would-be resummed quantum action would retain the property  of  (\ref {ac1})  which made a CC unphysical. 

\section{Beyond the Unconventional Path}

A proposal was put forward in \cite{Gabadadze:2014rwa} that appears to retain 
the good part of Tseytlin's approach, but gets rid of the difficulty with the gravity loops. 
This is achieved by introducing a second universe, endowed with its own 
metric $f$, that interacts with our universe only globally via the following action:
 \begin{equation}\label{acmain}
\mathcal{A}={V_f}{\bar S}+S_f. 
\end{equation}
Here, ${\bar S}$ is defined in (\ref{ac1}), and is  a functional of  the metric $g$ 
and the SM fields.  The action $S_f$  contains the Einstein-Hilbert (EH) term for $f$,  
but  is independent of the metric of our universe,  $g$.   From the point of view of 
the $f$-universe,  the term   ${V_f}{\bar S}$ represents  an addition to  a CC in  
that universe, since $V_f$ is its  invariant volume.

Thus, it is straightforward to see  from (\ref {acmain}) that an arbitrary cosmological 
constant introduced in $S$ -- the numerator in (\ref{ac1}) --
becomes a cosmological constant of the $f$-universe. Therefore, there is no big CC 
problem in the $g$-universe -- the CC of the $g$-universe gets entirely absorbed into the 
$f$-universe \cite {Gabadadze:2014rwa}.

We note that the  action   (\ref {acmain})   can be readily used to obtain the 
Hamiltonians for  both $g$- and $f$- universes.
The unusual feature   of (\ref {acmain})  is  the multiplier 
$V_f/V_g$. The latter contains no derivatives of the metrics. 
Thus, this multiplier affects the determination of the  canonical momenta of the $g$-universe 
only by a factor of $V_f/V_g$.  This however does not represent 
an impediment  for writing  down the classical Hamiltonian of the theory. 

How about quantum corrections? From the arguments presented after (\ref {expS}), it should 
be evident  that the quantum loop  effects will be small as long as $V_g/V_f \lsim 1$.  Let us elaborate 
on this. The kernel of the path integral  for (\ref {acmain})  can be written as follows, 
\be
{\rm exp}\left ( {i {\cal A \over {\hbar}} } \right ) = \int d\lambda\, dq\,  {\rm exp} \left \{{i \over \hbar} \left [ 
\frac{1}{q}S+\lambda \left( \frac{V_g}{V_f}-q \right) + S_f \right ] \right \}\,.
\label{expA}
\ee
Then,  the new Planck constant for quantization of $g$-gravity  is $\hbar q$, which is now equal 
to $\hbar (V_g/V_f)$.  As long as there are self-consistent classical solutions  for $g$- and $f$- metrics  
such that   $V_g/V_f \ll 1$, the quantum loop corrections to the classical  action 
will be negligible.\footnote {One should also   worry about: (a) new nonlocal 
corrections that would arise in (\ref {acmain}),  and (b) new loop-generated  
terms  containing  polynomials of $\lambda$ and $q$ in (\ref  {expA}).  Neither (a) nor (b) 
will cause problems, as to be addressed in Section \ref{secloop}.} 

Hence, the main remaining task  is to arrange  for the  dynamics of the $f$-universe  
to satisfy  $V_g/V_f \ll 1$,  in a technically natural way.  This was achieved 
in  \cite {Gabadadze:2014rwa} by postulating that $S_f$ is the action of 4D Einstein's
gravity with  a  negative cosmological constant,  which is somewhat larger  than the CC generated in the 
$g$-universe.  If so, then the CC of the $g$-universe gets entirely absorbed into the $f$-universe, and since its 
magnitude is  smaller than  that of the CC in the $f$-universe, 
it only modifies the $f$-metric  slightly.  As a result, there is a  classical solution  for which the metric of the 
$f$-universe is AdS$_4$, while the metric of the $g$-universe is flat!

However, we cannot stop here since  we need to describe the cosmic acceleration.  
This should be done by means other than using a CC. There could be a few options 
here.\footnote{For instance, a 4D quintessence model can be used to get dark energy
 in the framework described in Sections 3 and 4 of the present paper.} We choose 
to introduce dark energy  via massive gravity \cite{deRham:2010ik,deRham:2010kj}, or its 
extensions.\footnote{See reviews \cite {Kurt,Claudia,Fred} on theory and 
phenomenology of massive gravity and its extensions.} This theory  postulates a small graviton mass,  $m\sim 10^{-33}$ eV. Albeit the small parameter put in by hand, this is a technically natural way of introducing dark energy, since the graviton mass, unlike a CC, does not receive the additive power-divergent quantum corrections, and its renormalizations 
are multiplicative \cite {dRGHP}. 
 
Once the dark energy is introduced, the $g$-metric would turn into that of a dS$_4$ spacetime 
with a tiny observable curvature. Our mechanism will then be at work since the invariant volume of the  
universal cover of AdS$_4$  is infinitely larger than the infinite  
volume of dS$_4$. Thus, we have the desired result,  $V_g/V_f\rightarrow 0$,
and the classical solutions described above remain self-consistent even after the quantum 
gravity loops are taken into account \cite{Gabadadze:2014rwa}. 

We note that the above proposal  does not eliminate  the  motivation for quantum gravity. 
The latter  would be needed to  obtain the action ${\cal A}$ from  more fundamental theory,  and also to  
quantize the action $S_f$ in (\ref {acmain}).  Furthermore, the mechanism  does not  exclude  
higher powers of the curvature terms  and their derivatives in $S$, if they arise due to the 
$\alpha^\prime$ expansion of a putative completion into string theory. In fact, an infinite number of such  terms  
in $S$ would  be welcome  in  the present framework -- the black hole and cosmological 
singularities  could hopefully be resolved by these $\alpha^\prime$ 
terms (see more in Section \ref{secloop}).

\section{A New Embedding}

The construction of \cite {Gabadadze:2014rwa}  invoked  two  assumptions: 
(I) the CC in the $f$-universe was assumed to be  large and negative, without 
tunings of any kind;  (II) dark energy in the $g$-universe  was postulated to be 
due to massive gravity.  Both assumptions seem to be of  
a provisional character, as one could come up with other  scenarios 
giving $V_g/V_f \ll 1$, with dark energy produced by a different technically 
natural mechanism.

Having said this,  however,  we note that assumption (I) may have a strong justification:  
the $f$-universe with its negative CC  could  be supersymmetric, and  broken supersymmetry  
in the $g$-universe needs not be communicated to the $f$-universe,  
since $f$ and $g$  are only globally connected \cite {Gabadadze:2014rwa}.

How about assumption (II)? It is not necessary, but could there 
be advantages in using massive gravity, as opposed to other 
forms of dark energy (CC excluded)? Are there specific features of massive gravity 
that naturally fit into the above proposal for solving the big CC problem?  If the answers to these 
questions were positive, then the proposed mechanism   for the CC problem  would be connected 
to the mechanism for dark energy.  

In the present  work, we claim partial success in this quest. In particular, we will show  
how the fiducial metric written  by means of the St\"uckelberg fields -- that are necessary for  
the diffeomorphism  invariant  formulation   of massive gravity --   can be related 
to the embedding and metric of the 5D $f$-universe, introduced to solve the big CC problem. 
Our construction unveils a higher dimensional origin of the fiducial metric 
in massive gravity, and gives more appeal  to the scenario in which dark energy is due to 
massive gravity  and its extensions. 

As a by-product, we  will show that embedding of massive gravity 
into the action $S$ in (\ref {acmain})  helps deal with  a well-known  
difficulty of this theory  related to its  quantum strong coupling behavior 
at the scale  $\Lambda_3 =(M_{\rm Pl} m^2)^{1/3}$: the quantum loop corrections to 
the massive gravity action (and those in its extensions) are now controlled by 
$\hbar q \to 0$.  Therefore,  the  loop diagrams do not generate any counterterms, 
beyond the terms already present in the tree-level  action.    This 
does not eliminate  the  scale $\Lambda_3$,  and the theory is still 
strongly coupled at that scale.   However, the effective action receives no 
loop-generated counterterms.\footnote{For  detailed discussions  
of the  loops and optical theorem in this framework, see Section 6.}  
Thus,  the full quantum effective action has as few terms as the 
tree-level action, and calculations done in this theory -- albeit strongly coupled at $\Lambda_3$ -- 
are  exact, modulo the high derivative curvature terms that  presumably appear at the Planck scale 
($M_{\rm Pl} \ggg \Lambda_3$) as part of the completion at that scale.  This is unlike  a generic 
non-renormalizable  theory  that would  generate an infinite number of  new  
counterterms at the scale $\Lambda_3$.

The  main idea  of the present work is to regard   the action of the ``other" universe as the EH 
action of  5D gravity,  described by the metric $F$ on  AdS$_5$. Hence, we use 
the notation  $S_F$,  instead of $S_f$  used previously,  to indicate that the ``other" universe 
is five-dimensional; we denote its invariant volume by $V_F$. We then endow a hypersurface at the AdS$_5$ boundary  with the dynamical metric $g$,  and  
couple $g$ to the pullback of the bulk metric,  $\gamma$,  in such a way  that the latter acts as the fiducial metric of 4D massive gravity described  by $g$.

The pullback $\gamma $ can be expressed via  the  four St\"uckelberg  
scalars of massive gravity $\varphi^a(x)$,  ${ \gamma}_{\mu\nu} = \partial_\mu \varphi^a 
\partial_\nu \varphi^a  \eta_{ab} ~(a,b,... =0,1,2,3)$. These scalars  
then end up parametrizing the 4D hypersurface at the  boundary of the AdS$_5$ bulk.
The detailed construction is worked out in  Section \ref{secm1}, where a simple local 
model without the $V_F/V_g$ factor is considered.  Besides its illustrative purpose, this model also 
delineates a method of introducing gravity, albeit massive, on the boundary of AdS$_5$. 
The full model (\ref{acmain}) is examined in Section \ref{secm2}, where we show 
how the fiducial metric of massive gravity emerges from the pullback 
of the bulk metric in AdS$_5$. The discussions of the quantum loop effects and the 
strong coupling problem of massive gravity in this approach are given in Sections \ref{secloop} and 
\ref{secstrong}, respectively. Section \ref{seclast} contains comments and outlook.

\section{Massive Gravity On Top of the Boundary of AdS$_5$}\label{secm1}

Our bottom-up construction  begins with postulating a dynamical 4D metric,  $g_{\mu\nu}$, 
on top of the boundary of AdS$_5$. The metric  $g$  is not related to  a pullback of the bulk gravity.
Hence,  such a  boundary has more structure than  
the conventional conformal boundary of AdS$_5$, and may be thought as 
some geometric hypersurface, that is 
endowed with  $g$ and  is placed right at the boundary  of AdS$_5$.

To set the conventions, $x^{\mu}$ (with $\mu, \nu,...=0,1,2,3$)  denote the  coordinates  of 
4D spacetime of the hypersurface  that is endowed with the  dynamical metric $g_{\mu\nu}(x)$.
The massive gravity  action  defined via $g$  on the world-volume of the boundary 
hypersurface reads as follows, 
\begin{equation}\label{mg1}
S={M^2_{\rm Pl}} \int d^4x \sqrt{|g|}\left[ R(g)-2\Lambda+2m^2\mathcal{U} \left({\cal K} \right) \right],
\end{equation}
where  the  diff-invariant potential  $\mathcal{U}$  was built in \cite{deRham:2010ik,deRham:2010kj}, 
as a function of the inverse metric $g^{-1}$,  and the 
fiducial Minkowski metric, ${ \gamma}_{\mu\nu} = \partial_\mu \varphi^a 
\partial_\nu \varphi^a  \eta_{ab} ~(a,b,... =0,1,2,3)$, in an arbitrary 
coordinate system;    this potential  can be written in  the following form:
\begin{equation}\label{mGRUK0}
\mathcal{U}(\mathcal{K})=\det_2(\mathcal{K})+\alpha_3\det_3(\mathcal{K})+\alpha_4\det_4(\mathcal{K}),
\end{equation}  
where  the matrix $\mathcal{K}=1-A$, with $A$ being  defined  as one of the roots of,  
$A^{\mu}_{\alpha}A^{\alpha}_{\nu}=g^{\mu \alpha}\gamma_{\alpha \nu}$, so that 
$\mathcal{K}=1-\sqrt{g^{-1}\gamma}$ \cite{deRham:2010kj}.

The fiducial Minkowski metric, $\gamma$, is not dynamical,  and its  origin is unknown in massive gravity,  
because the dynamical origin of the fields $\varphi^a$  is not known.   In the high energy limit, these 
fields parametrize the  helicity $\pm 1$ and helicity $0$ degrees of freedom of a massive graviton.  
Geometrically, they  can be regarded as coordinates  of a certain fiducial  4D Minkowski space, that 
is postulated as "pre-geometry" in pure massive gravity (see discussions in \cite {GGKurt}).\footnote{Note that the origin of the fields $\varphi^a$ is not elucidated  
in conventional 4D  bigravity theories either for the following reasons: 
these theories  have two dynamical metrics, $g (x)$ and 
$\tilde \gamma(x)$,  however only one  common diffeomorphism for  the two; 
to restore   the second diffeomorphism in bigravity,  one has to introduce the  four fields,   $\varphi^a$, 
and postulate that  the second metric  is  a function of these fields,  $\tilde \gamma (\varphi )$.   While the  diff  
invariance, as any local gauge symmetry,  is a redundancy of description, 
it is a helpful  redundancy, and in the case  of 
massive gravity and bigravity,  it calls for understanding  of 
the  dynamical origin of the  helicity $\pm 1$ and helicity $0$ degrees 
of freedom that are encoded in $\varphi^a$.  
The lack of this understanding  is the reason why both massive gravity and bigravity in the existing formulation 
are strongly coupled theories  at an energy scale that is much lower than  
the Planck scale.  The above comments   can be briefly 
summarized as follows: the analog of the  Lorentz-invariant Higgs mechanism for  
gravity  or bigravity -- from which $\varphi^a$'s  would originate  -- is not known at present.

Before we proceed, we make a
few comments on the literature:  A partial list of the earlier bigravity theories   is in 
\cite {bigravity1,bigravity2}. That the problem, now referred as  the Boulware-Deser  (BD) ghost  problem,  
potentially applies to all massive gravity and bigravity theories  was shown  in Ref. \cite {BD} .  
After  the terms (\ref {mGRUK0}) were proposed in Ref. \cite{deRham:2010kj} as ghost free terms, proven to 
be free of the BD ghost up to the fourth order in nonlinearities \cite{deRham:2010kj},  and proven  to be BD 
ghost free to all orders in \cite {Hassan:2011hr}, these terms  were invoked by Ref. \cite {exten2}   into  the  
earlier bigravity theories  \cite {bigravity2}  and  it was shown   that the resulting theories are free of the 
BD ghost  \cite {exten2}.  This prove, and the results of works  \cite{KurtRachel}  and 
\cite {Deffayet} that reformulated  these theories  in the  vierbein formalism,  
imply  that the earlier versions  of bigravity \cite {bigravity1,bigravity2} -- except two bigravity 
models  by Wess and Zumino (WZ) in \cite {bigravity1}  --  suffer  from the  BD ghost. 
Two out of the four WZ models introduced the subset of terms (\ref {mGRUK0}),  
written in the first order formalism for two tetrads, albeit the WZ work does not address the 
nonlinear consistency and the BD problem, since it precedes the BD work by two years or so.} 

While this work does not provide the Higgs mechanism for gravity, 
it supplies geometric  meaning  for  the fields $\varphi^a$. 
This is done  in the present approach  by relating 
the fiducial metric to  the pullback of  the  bulk metric, where $\varphi^a$'s get related to 
5D coordinates. To see this relation we denote  the bulk coordinates in AdS$_5$ as  
$Y^A$'s with $A,B,...=0,1,2,3,5$, and relate the  5D and 4D coordinates as follows, 
\begin{equation}
Y^{A}=\{\varphi^a(x),z \}.
\end{equation}
In this coordinate system,   the 4D boundary hypersurface is located at $z=0$, 
while the fields  $\varphi^a(x)$ (with $a,b,...=0,1,2,3$) give 
different  parametrizations  of the hypersurface. In other words, these fields 
map the 4D spacetime  on which the $g(x)$ metric lives,  onto the  
4D hypersurface in  5D AdS spacetime.  

  As to the bulk action,  it is the 5D EH action with the negative CC and 
the Gibbons-Hawking  term and boundary counterterm included (see, e.g., \cite{Liu:1998bu}),
\begin{equation}\label{s5}
S_F=\int_{M} d^5Y~\sqrt{|F|}\left(R(F)+\frac{12}{l^2} \right)+2\int_{\partial M}d^4Y~\sqrt{|{\tilde \gamma}|}K-\frac{6}{l}\int_{\partial M}d^4Y~\sqrt{|{\tilde \gamma}|}.
\end{equation}
Here, the 5D Planck mass $M_5$ is set to unity, $\tilde \gamma$ is the induced metric on the boundary, 
and $K$ is the trace of the boundary extrinsic curvature.  Our ultimate goal is to study the functional 
\be \label{SSF}
{\cal A}_5 = {V_F\over V_g} \, S \,+\,S_F,
\ee
but in this section, as a warm-up exercise,  we study  (\ref {SSF}) without 
the $V_F/V_g$ factor in front of $S$. Thus, the total action of  the  theory  considered in 
this section is 
\begin{equation}\label{acm1}
S_{\rm tot}=S+S_F.
\end{equation}
The action (\ref {s5})  contains a negative cosmological constant, 
$-\frac{6}{l^2}$, and thus generates an $\text{AdS}_5$ solution to the respective Einstein equations,
\begin{align}\label{m1}
ds^2&=F^{(0)}_{AB}dY^AdY^B \nonumber \\ &=\frac{l^2}{z^2}\left(\eta_{ab}d\varphi^{a}
d\varphi^{b}+dz^2 \right) \\ &=\frac{l^2}{z^2}\left(\eta_{ab}\partial_{\mu}\varphi^{a}
\partial_{\nu}\varphi^{b}dx^{\mu}dx^{\nu}+dz^2 \right). \nonumber
\end{align}
The interval in the last line  is written in terms of the boundary massive gravity coordinates. 

We are now in a position to relate the induced  metric on the boundary, $\tilde{\gamma}$, to  
the fiducial metric of massive gravity, $\gamma$.   Our choice for this relation reads as
\begin{equation}\label{bm1}
\gamma _{\mu \nu}(x)=\lim_{z\rightarrow 0} {z^2\over l^2}{\tilde \gamma}_{ab}(z,\varphi(x))
\frac{\partial \varphi^a}{\partial x^{\mu}}\frac{\partial \varphi^b}{\partial x^{\nu}}.
\end{equation}
This  identifies the St\"ukelberg fields used in the  massive gravity potential as the coordinates of the 
boundary in the AdS$_5$ bulk. 
The variational procedure for the total action is defined as follows:  $S_F$ is varied w.r.t. $F$. 
The boundary action, $S$,  on the other hand,  is  varied w.r.t. $g$, but no variation is taken w.r.t. $F$.
The induced bulk  metric on the boundary, $\tilde \gamma$,  and the fiducial metric, $\gamma$,
are then related to each other via the boundary condition  (\ref {bm1}). This procedure is compatible with the variational principle in AdS/CFT, since varying the boundary fields does not affect the bulk dynamics.  
Under this premise, the linearized theory with Dirichlet boundary data is studied 
next.\footnote{We use $S_F$ as the bulk action even though it contains the  Gibbons-Hawking 
boundary term and the classical counter-term, the  role of which are to guarantee  the right bulk  
Einstein equations.   In contrast,  the action $S$, referred here as the boundary action, 
gives rise to the dynamical equations of motion on the boundary.  We hope this nomenclature 
will not cause any serious confusion.}

The derivation of the on-shell boundary effective action  from  the bulk action, $S_F$, 
is a well-known calculation in standard AdS/CFT. Here we give an outline of this procedure 
adapted to our action (\ref{acm1}). For details, we refer the readers to Appendix \ref{append1}.

In the following, we will  work at the regularized boundary $z=\epsilon$,  and  the limit $\epsilon\rightarrow 0$  will be 
taken only at the end of the calculations. In the linearized theory, the full bulk metric is $F_{AB}= F_{AB}^{(0)}+h_{AB}$, where $h_{AB}$ denote the components of 5D metric fluctuations. We work in the gauge $h_{0B}=0$ and introduce $h_{ab}(z,\varphi)=z^2h_{AB}\delta^A_a\delta^B_b/l^2$. Then we can  write
\begin{equation}
ds^2=\frac{l^2}{z^2}\left[(\eta_{ab}+h_{ab})d\varphi^{a}d\varphi^{b}+dz^2 \right].
\end{equation}
By (\ref{bm1}), the fiducial metric on the boundary receives corrections from the 5D fluctuations,
\begin{equation}
\gamma_{\mu \nu}(x)=\eta_{ab}\partial_{\mu}\varphi^a\partial_{\nu}\varphi^b+\delta\!\gamma_{\mu \nu}(x),
\end{equation}
 where
\begin{equation}
\delta\!\gamma_{\mu \nu}(x)=\lim_{\epsilon\rightarrow 0} h_{ab}\left(\epsilon,\varphi(x)\right)\partial_{\mu}\varphi^a \partial_{\nu}\varphi^b.
\end{equation}
Thus, $\delta\!\gamma_{\mu \nu}$ represent the induced fluctuations. We also define the related metric $\gamma_{ab}(\varphi(x))=\eta_{ab}+h_{ab}(\epsilon,\varphi(x))|_{{\epsilon\rightarrow 0}}$, so that $\gamma_{\mu \nu}(x)=\gamma_{ab}(\varphi(x))\partial_{\mu}\varphi^a \partial_{\nu}\varphi^b$. The resulting  effective boundary 
action can be written as a sum of various terms
\begin{equation}
S[h_{ab}]=S_{\partial^2}+S_{\partial^4}+S_{\text{nonlocal}},
\end{equation}
where the respective parts of the total action are defined as follows:
\begin{equation}\label{dd2}
S_{\partial^2}\sim ~\frac{1}{\epsilon^2}\int d^4\varrho ~h\partial^2h,\quad S_{\partial^4}\sim ~\ln \epsilon \int d^4\varrho~h\partial^4h,
\end{equation}
\begin{equation}
 S_{\text{nonlocal}}\sim \int d^4\varrho d^4\varsigma~\frac{h(\epsilon,\varrho)h(\epsilon,\varsigma)}{|\varrho-\varsigma|^8}.
\end{equation}
The exact formulae and tensor structures can be found in (\ref{d2}) -- (\ref{nl1}). In particular, the nonlocal terms give the two-point function for the boundary CFT stress-tensor,

\begin{equation}
S_{\text{nonlocal}}=-\frac{1}{2}\int d^4\varrho d^4\varsigma ~h_{ab}(\epsilon,\varrho)h_{cd}(\epsilon,\varsigma)\Expect{\T^{ab}(\varrho)\T^{cd}(\varsigma)},
\end{equation}
\begin{align}
\Expect{\T^{ab}(\varrho)\T^{cd}(\varsigma)}\equiv~&\frac{20}{\pi^2}\frac{1}{|\varrho-\varsigma|^8} \nonumber \\
&\times \left[\frac{1}{2} J^{ad}(\varrho-\varsigma)J^{bc}(\varrho-\varsigma)+\frac{1}{2}J^{ac}(\varrho-\varsigma)J^{bd}(\varrho-\varsigma)-\frac{1}{4}\eta^{ab}\eta^{cd} \right], \nonumber
\\ J^{ab}(\varrho)=&\eta^{ab}-2\frac{\varrho^a \varrho^b}{|\varrho|^2}.
\end{align}
As a consistency check, the $S_{\partial^2}$ terms comprise the standard kinetic terms for linearized Einstein gravity,
\begin{equation}\label{kin1}
S_{\text{kin}}[h_{ab}]=\frac{1}{2\epsilon^2}\int d^4\varrho~\left[\frac{1}{4}h_{ab}\square h^{ab}-\frac{1}{2}h_{ab}\partial^a\partial_ch^{bc}-\frac{1}{4}h\square h+\frac{1}{2}h_{ab}\partial^a\partial^b h \right].
\end{equation}

The local terms in (\ref{dd2}) with divergent powers of $\epsilon$ are cancelled by appropriate counter-terms introduced as part of the renormalization procedure. This procedure deprives the fields $h_{ab}(\epsilon,\varphi)$ of any dynamics. As a result, one 
regards  them as the Dirichlet  data supplied by boundary sources. 
We highlight this fact by using the notation  $h^0_{ab}(\varphi(x))=h_{ab}(\epsilon,\varphi(x))|_{\epsilon\rightarrow 0}$, so that $\gamma_{ab}(\varphi(x))=\eta_{ab}+h^0_{ab}(\varphi(x))$. Then the total effective boundary action is
\begin{align}\label{mGRunmod}
S^{\rm eff}_{\text{tot}}[g,\gamma]=&{M_{\rm Pl}^2}\int d^4x\sqrt{|g|}\left[ R(g)-2\Lambda+2m^2\mathcal{U}\left( {\cal K} \right) \right] \nonumber \\
&-\frac{1}{2}\int d^4\varrho d^4\varsigma~h^0_{ab}(\varrho)h^0_{cd}(\varsigma)\Expect{\T^{ab}(\varrho)\T^{cd}(\varsigma)}+ {\cal O} \left (  (h^0)^3 \right )\,,
\end{align} 
where ${\cal K} =  1 -\sqrt{g^{-1}\gamma_{ab}(\varphi(x))\partial_{\mu}\varphi^a \partial_{\nu}\varphi^b}$, 
contains a general  fiducial metric \cite{exten2, exten1}; formally, the theory with $h^0\neq 0$
differs from massive gravity with a general fiducial metric discussed in  \cite{exten2, exten1} by an infinite 
number  of new polynomial terms  in $h^0$, and can be regarded as a theory in  
some external  background field  set by $h^0$.  However, these new polynomial terms  do not enter   the massive 
gravity equations of motion since they do not depend on either $g$ or $\varphi^a$.  
Hence, when  viewed as a 4D theory,  the obtained model is nothing but  massive gravity 
with an arbitrary fiducial metric parametrized by $h^0$ \cite{exten2, exten1}. However, 
this is a limited view since the terms containing 
higher powers in  $h_0$  encode   additional useful information:
$h^0$  can be regarded as as external source
for a dual  4D  field  theory of the 5D AdS gravity, and if so, 
the  polynomials in $h^0$ define dual CFT correlators: $n$th variation 
of the action (\ref {mGRunmod}) w.r.t. $h^0$, with the subsequent substitution $h^0=0$, 
calculates the CFT $n$-point correlation function,  modified by the terms that are proportional to powers of the 
graviton mass. Thus,  for a small graviton mass,  one would arrive at a softly broken CFT.
To summarize, our prescription is to put $h^0 =0$ after all the calculations are done.
On the gravity side of the dual pair,   this  condition is  a choice of the boundary values.
On the CFT side, this choice -- imposed after all the variations are done --  
enables us to get  the CFT  correlation functions.  

Likewise, the corrections to the fiducial metric can be identified, up to diffeomorphisms,  
with the Dirichlet boundary data,
\begin{equation}
\delta\!\gamma_{\mu \nu}(x) =h^0_{ab}(\varphi(x))\partial_{\mu}\varphi^a \partial_{\nu}\varphi^b.
\end{equation}
Thus, both the CFT and the massive gravity sectors are sourced by the same boundary fields. If the 
standard AdS/CFT prescription is applied to the above action, (\ref {mGRunmod}), the CFT correlation 
functions will be modified by terms that are proportional to powers of the graviton mass. Thus,  for a small graviton mass, 
one would arrive at a softly broken CFT, but this is not our research topic for the time being.

\section{Removing CC and Introducing  Dark Energy}\label{secm2}

We now turn to the action functional that in the present context 
can remove an arbitrary 4D CC from the  dynamical 4D boundary of 5D AdS gravity. 
In the spirit of (\ref{acmain}), we look at the action (\ref{SSF}),
\begin{equation}\label{modaction}
{\cal A}_5=\frac{V_F}{V_g}S+S_F.
\end{equation}  
Here, $V_F=\int d^5Y \sqrt{|F|}$ and $V_g=\int d^4x\sqrt{|g|}$. All the conventions are the same as in 
the previous sections. To avoid ambiguity in the equations of motion due to ratios of infinite volumes, we regularize all integrals in (\ref{modaction}),
\begin{align}
\int d^4x &= \lim_{\bar{x}^0,...,\bar{x}^3\rightarrow \infty} \left(\prod_{\mu=0}^3 \int_{-\bar{x}^{\mu}}^{\bar{x}^{\mu}} \right) d^4x \equiv \lim_{\bar{x}\rightarrow \infty}\int_{\text{reg}}d^4x ,\nonumber \\
\int d^5Y &=\int dz~d^4\varphi= \lim_{\bar{\varphi}^0,...,\bar{\varphi}^3\rightarrow \infty} \lim_{\epsilon \rightarrow 0}  \int_{\epsilon}^{\infty}dz \left( \prod_{A=0}^3 \int_{-\bar{\varphi}^A}^{\bar{\varphi}^A} \right) d^4\varphi~ \equiv \lim_{\bar{Y}\rightarrow \infty} \lim_{\epsilon \rightarrow 0} \int_{\text{reg}}d^5Y,
\end{align}
so (\ref{modaction}) is actually
\begin{equation}
{\cal A}_{5}=\lim_{\bar{x},\bar{Y}\rightarrow \infty } \lim_{\epsilon \rightarrow 0} \left[ \frac{V_F}{V_g}S+S_F\right]_{\text{reg}}.
\end{equation}
The order of  the limits is carefully arranged: $\epsilon\rightarrow 0$ is taken before $\bar{x},\bar{Y}\rightarrow \infty$. This regularization first restricts the theory to a 5D ``box," then extends the $z$ direction to include the conformal boundary, before taking  the remaining  limits. For conciseness, we 
will always treat the volumes as regularized, but will not write the  regularizations explicitly.

It is clear that in the action (\ref {modaction}), the 4D CC does not curve the 4D spacetime, 
but instead gets absorbed into the 5D CC. If the magnitude of the 5D CC is greater 
than that of the 4D CC, then this leads to  an obvious modification of the curvature of the 
AdS bulk, while all the conclusions of the previous section remain valid.  An order-of-magnitude  hierarchy between
the scales of the 5D and 4D CC's is straightforward to arrange, since  5D theory can be 
exactly supersymmetric, while the broken SUSY in 4D could guarantee its scale to be lower.\footnote{For similar arguments, 
see \cite{Gabadadze:2014rwa}.} Moreover, since $V_F$ now is the five-volume of 
the 5D AdS spacetime, and $V_g$ is going to be  the  four-volume 
of the 4D dS spacetime, the ratio $V_g/V_F \to 0$,  in Planck units. This guarantees 
that the outlined classical solution is stable w.r.t. quantum loops.

What remains to be seen is how the fiducial metric 
arises in  the boundary massive gravity, and how the theory gives rise to dark energy. 
To this end, we just repeat, step by step, the procedure of ``integrating out" the AdS 
bulk described in the preceding section. The result,  in the quadratic order in the bulk,  
reads as follows:  
\begin{align}\label{btension}
S^{\rm eff}_{\text{tot}}[g,\gamma]=& \frac{V_F}{V_g}{M_{\rm Pl}^2}\int d^4x\sqrt{|g|}\left[ R(g)+2m^2\mathcal{U}\left( {\cal K} \right) \right] \nonumber \\
&-\frac{1}{2}\int d^4\varrho d^4\varsigma~h^0_{ab}(\varrho)h^0_{cd}(\varsigma)\Expect{\T^{ab}(\varrho)\T^{cd}(\varsigma)}+ {\cal O} \left (  (h^0)^3 \right )\,.
\end{align}
The potential  ${\cal U}$  appearing in the above action was defined in (\ref {mGRUK0}), and, as before, 
${\cal K} =  1 -\sqrt{g^{-1}\gamma_{ab}(\varphi(x))\partial_{\mu}\varphi^a \partial_{\nu}\varphi^b}$. Both the CFT and the massive gravity sectors essentially stay the same (for further comments, see Section \ref{seclast}), despite the global modification. The difference is that the 4D cosmological constant, $\Lambda$, is removed from the $g$ dynamics, and the volume factor $V_F$ is taken on  the bulk solution for $F$.  

To make things clearer, we emphasize a subtlety in  how the effective boundary action   
(\ref{btension}) is obtained.  Upon varying $\mathcal{A}_5$ w.r.t. $F$ the 
equations of motion are,  $G_{AB}=(\tilde{\Lambda}-\bar{S}/2)F_{AB}$, where $\tilde{\Lambda}$ is a constant, 
${\bar S}$ is a constant-valued functional, 
and $G_{AB}$ is the Einstein tensor for $F$. 
Thus $\bar{S}$ contributes to the bulk CC,  and this renormalizes  the CC of the $F$-universe, 
as discussed  in detail in \cite {Gabadadze:2014rwa}. We've already moved the 
4D CC from the $g$-universe into the  $F$-universe, but the $g$-universe can also produce 
a  self-accelerated background with curvature $\sim m^2$, that would also contribute
to the bulk CC.\footnote{ We note however, that the latter is  many orders of magnitude smaller  
compared to a typical  curvature in the $F$-universe \cite {Gabadadze:2014rwa}, 
and can be ignored for all ``practical purposes" of the bulk physics as we've done above. 
However, one could  be more general and retain this addition to the bulk CC. In that case
the boundary counterterm eliminating the bulk-induced  classical divergencies should also 
be adjusted accordingly. This would not change our result for the remaining 4D effective action.}
Then,  one integrates out $F$,  as in AdS/CFT, and thus one naively seems to be able  to  
reduce  $\mathcal{A}_5$ to a boundary action dual to a pure, unbroken CFT corresponding to the second line of (\ref{btension}). However, this is false: there ought to be an additional contribution to the boundary effective action due to the 
$g$-metric.   To see this, we note that variation of $\mathcal{A}_5$ w.r.t. 
$g$ gives rise  to 4D equations of motion  for dynamical gravity described by the metric $g$.  
In order for the effective boundary action (\ref{btension}) to capture this, it has to include 
the first line. As a result, the $g$-metric  couples to a fiducial metric, $\eta+ h^0$,  obtained from the bulk metric
and boundary conditions.  
The fiducial metric  at this point is not yet specified and the theory looks like  bigravity   \cite {exten2}, 
but with a different action for $h^0$ given  in (\ref{btension}) in the lowest order in $h^0$. 
As we've already noted at the end of Section 4, the CFT correlators obtained by 
varying  (\ref{btension}) w.r.t. $h^0$, and subsequently putting $h^0=0$, will contain  local 
pieces proportional to the graviton mass.  Thus  CFT correlation functions are  amended by local terms 
proportional to the graviton mass. On the other hand,  the resulting  gravity equations will  be those of massive gravity 
with Minkowski fiducial metric  \cite{deRham:2010ik,deRham:2010kj}, since $h^0=0$ should be 
used in the end to comply with  the AdS/CFT prescription.

From (\ref{btension}), the relevant part of the action is
\begin{equation}\label{mGRmod}
S_g=\frac{V_{F}}{V_g} M_{\rm Pl}^2\int d^4x \sqrt{|g|}\left[ R(g)+2m^2\mathcal{U} \left({\cal K} \right) \right],
\end{equation}
with the fiducial metric $\gamma_{\mu \nu}=\left[\eta_{ab}+h^0_{ab}(\varphi(x)) \right]\partial_{\mu}\varphi^a\partial_{\nu}\varphi^b$, determined  by the Dirichlet boundary data $h^0_{ab}$, and 
the volume factor $V_F$  taken on  the bulk solution for $F$. 
 Setting $M_{\rm Pl}$ to 1, the $g$-metric Einstein equations can then be written as nine traceless and one trace equations \cite{Gabadadze:2014rwa},
\begin{equation}
R_{\mu \nu}-\frac{1}{4}g_{\mu \nu}R=T_{\mu \nu}-\frac{1}{4}g_{\mu \nu}T,
\end{equation}
\begin{equation}\label{eom22}
R+T= \langle T \rangle -2m^2 \left\langle g^{\mu \nu} \frac{\partial \mathcal{U}}{\partial g^{\mu \nu}} \right\rangle, 
\end{equation}
\begin{equation}
T_{\mu \nu}=\frac{-2}{\sqrt{|g|}}\frac{\delta }{\delta g^{\mu \nu}}\int d^4x \sqrt{|g|}m^2\mathcal{U}=m^2\left( g_{\mu \nu}\mathcal{U}-2\frac{\partial \mathcal{U}}{\partial g^{\mu \nu}} \right),
\end{equation}
where $\langle \cdots \rangle$  denotes the spacetime average,
\begin{equation}
\langle \cdots \rangle\equiv\frac{\int d^4x \sqrt{|g|} (\cdots)}{V_g}.
\end{equation}

Note that on self-accelerated solutions, $R,T \sim m^2$ are equal to constants, thus $R=\langle R \rangle$ and $T=\langle T \rangle$ should be used in (\ref{eom22}). Also, $g^{\mu \nu} \partial \mathcal{U} / \partial g^{\mu \nu}=C(\alpha_3,\alpha_4)$ is a constant on the self-accelerated solutions, with $\alpha_3$ and $\alpha_4$ being the free parameters in (\ref{mGRUK0}). Its spacetime average yields the same constant, and therefore (\ref{eom22}) is reduced to
\begin{equation}
R=-2m^2C(\alpha_3, \alpha_4).
\end{equation}  
Hence, for some reasonable choices of parameters, we may get $m^2\sim H_0^2$ in a technically natural way, with $H_0$ being the Hubble constant.\footnote{Note that in massive gravity one would have gotten a similar equation, $ R=-2m^2{\tilde C}(\alpha_3, \alpha_4) $, but with a different function of the parameters, $\tilde C$;
this is because  equation (\ref {eom22}) differs from the corresponding massive gravity equation, $R+T=0$. How this 
difference affects the faith of fluctuations on the self-accelerated solution needs to be investigated.} This conclusion is not changed by introducing quasidilaton \cite{quasidila}, which only affects and improves the dynamics of small perturbations by removing unstable and superluminal modes \cite{Gabadadze:2014kaa}. Furthermore, instead of massive gravity or quasidilaton, one can 
straightforwardly use any of their known extensions, see, e.g., \cite {exten2,exten1,extensions} -- the key point is 
for the theory to have a fiducial metric and for the dark energy to be given by the stress-energy tensor associated with 
graviton mass.  As such, our approach establishes a theory on the AdS$_5$ boundary that removes the big cosmological constant and generates a small curvature given by the graviton mass. Meanwhile, it also ascribes natural meanings to the fiducial metric and the diffeomorphism St\"ukelberg fields in the massive gravity.

 \section{The Loop Expansion}\label{secloop}

In this section, we briefly review the quantization algorithm in which 
the SM fields are quantized with the 
Planck constant $\hbar$, whereas gravity with the rescaled Planck constant, 
$\hbar q$, where $q$ is a functional whose magnitude is determined 
by classical equations of motion  \cite {Gabadadze:2014rwa}. 
In particular, the classical solutions that we consider  as relevant for our purposes
all give  $q\to 0$. Therefore, our theory reduces to one in  which  all the fields 
but  gravity are  quantized in a conventional manner, while  
gravity is kept classical,  al least at low energies, since $\hbar q \to 0$. 

Proposals to quantize the SM,  and couple it to classical gravity, have been extensively discussed
in the past,  see  works \cite {ClassGravity,InIn,Anselmi}  and  references therein. 
That  such a scheme should exist at least as an approximation -- to describe 
our empirical experience  with quantized SM  and classical  gravity so far observed  in our universe --
is undeniable. More subtle is the question that whether  the observed 
gravity in our universe can be classical as a matter of principle,  and be consistently 
coupled to quantized SM; this is the question addressed affirmatively  in \cite {ClassGravity,InIn,Anselmi} 
and references therein. While in our approach we will not need to  
assume  that a fundamental theory of gravity is not quantized, we will nevertheless 
consider a case where at low energies the effective Planck's constant 
for gravity, $\hbar q$,  tends to zero, and in that respect, the  present section describes   
a straightforward  application  of some of the techniques   of \cite {ClassGravity,InIn,Anselmi}  
to our approach.  In particular,  we will use the method of Ref. \cite {Anselmi} (we could 
use equally well  the in-in formalism and its earlier versions  reviewed  in \cite {InIn}).  
While the approach  of Ref. \cite {Anselmi} corresponds to our  case $\hbar q=0$, we 
argue that the limiting procedure,  $\hbar q \to 0$,  can also be well defined and meaningful. 
 
We then apply this algorithm to massive gravity, and 
argue that this procedure simply removes the problem of otherwise 
unmanageable counterterms  at the strong coupling scale of that theory. 

\subsection{Non-gravitational Loops}

We assume that classical gravity should be completed into a full-fledged 
quantum  theory with good UV behavior at  a certain  energy scale, $M_{\rm QG}$
(the Planck scale,  or string scale),   that is higher, by  at least an order of magnitude,  
than  the particle physics UV scale, $M_{\rm SM}$.  The latter is assumed to be a scale  
at which the particle physics interactions themselves (with gravity switched off) become UV complete, for instance, 
in  an  asymptotically free grand unified theory (GUT). Then, it is not unnatural 
to  expect a  hierarchy between  $M_{\rm QG}$  and  $M_{\rm SM}$ of two orders of magnitude.  
For definiteness, we will assume that $M_{\rm QG}$ is of order the Planck scale, $M_{\rm QG}\sim 10^{18}~{\rm GeV}$,
while $M_{\rm SM}$ is of order the GUT scale, $M_{\rm SM}\sim 10^{16}~{\rm GeV}$, but our 
discussion does not really depend on these concrete values.

Owing to this hierarchy,  gravity should be  well approximated by a 
classical field theory  at and below  the energy scale $M_{\rm SM}$.  
In this low energy approximation,  the  path integral can be  defined  for all  
the SM fields  quantized with  ${\hbar}$,  while  gravity can be regarded 
as a classical field.   For self-consistency,    subsequent quantization of gravity should only give rise to 
negligible corrections to the quantized SM results. Since the corresponding formalism 
is well know (see  \cite {InIn,Anselmi} and references therein) 
our discussions in this section  will be brief and schematic. 

Thus, the Feynman  integral  for the quantized SM interactions at and below $M_{\rm SM}$ reads
\be
Z(g, J_n)  = {\rm const} \times \, \int d\mu ({\tilde \psi}_n) {\rm exp}\left ({i\over \hbar} \int d^4 x\sqrt{|g|} \left 
(  {\cal L} (g, {\tilde \psi}_n)   + J_n {\tilde \psi}_n  \right ) \right )\,,
\label{Z}
\ee 
where $d\mu({\tilde \psi}_n)$ is the  measure for  all the SM 
fields,  ${ \tilde \psi_n}$.\footnote{Gauge fixing and the Faddeev-Popov determinant  are included in this measure.}

As we have emphasized,  the metric  $g$ is regarded as an external field for now, 
and so are the sources,  $J_n$, introduced for every SM field. 
The 1PI effective action can be defined 
via  a Legendre transform of $ W(g, J_n) =- i \ln Z(g, J_n)$ as follows,
\be
\Gamma_{\rm {1PI}} \equiv   W (g, J_n) - \int d^4 x \sqrt {|g|} J_n \psi_n,  
\label{legandre}
\ee
where  $\sqrt{|g|} \psi_n  \equiv  -i  \delta \ln Z(g, J_n)/\delta J_n$ is  $\sqrt{|g|}$ times 
the expectation value of the SM field $\tilde \psi_n$  in the presence of the source $J_n$.  
These expectation values,  upon which the 1PI action depends, are referred to as ``classical fields," 
to emphasize that in the conventional approach they are not to be quantized further, since 
all the quantum corrections due to the SM interactions are
already taken into account in $\Gamma_{\rm {1PI}}$ and $\mathcal{L}_{\rm SM}$.

The  ``classical fields" $\Psi_n$, as well as the 1PI action, $\Gamma_{\rm {1PI}}$,  are not real 
in general even  for a subset of the original fields, ${\tilde \Psi}_n$,   that could be real.  
On the other hand, one does need a real  effective action  to couple to  gravity  in a conventional manner
in (\ref {ac1}). As  shown in Ref. \cite {Anselmi}  the real part of the 1PI action,
\be
\Gamma_{\rm eff} \equiv {\rm Re} \, \Gamma_{\rm {1PI}}\,\equiv \int d^4 x \sqrt {|g|} \mathcal{L}_{\rm SM}  (g,  \psi_n) ,
\label{Gammaeff}
\ee
can consistently be constructed  and used as a quantum  effective action  to which classical 
gravity can couple \cite {Anselmi}. Hence, the classical gravity equation would read as follows:
\be
R_{\mu\nu} -{1\over 2} g_{\mu\nu} R = - {2\over \sqrt{g}} {\delta \over \delta g^{\mu\nu}} \,
\Gamma_{\rm eff}\,.
\label{GR+SM} 
\ee
What follows in the next subsection is just a straightforward  application  of 
this procedure  of coupling  classical gravity to the quantized SM fields 
to  our case. It amounts to  an insertion of  the effective action
(\ref {Gammaeff}) into   (\ref {ac1}),  to account for  dynamical gravity. While for small but nonzero  
$q$ this procedure would require more careful study, for $q \to 0$ it 
reduces to the known one \cite {Anselmi};  this is enough for our purposes.\footnote{ While we use the 
technical tool developed  in  Ref. \cite {Anselmi}, we also  note here  a conceptual difference of our framework.  In the 
approach of   \cite {Anselmi} Einstein's gravity is considered to be a fundamental classical 
theory  and quantum matter fields are coupled to it. Due to  the matter loop corrections, 
one generates  the terms that are quadratic   in curvature invariants, $R_{\mu\nu}^2$.... 
No higher powers of curvature get generated since with the quadratic curvature 
counterterms the theory is shown to be renormalizable \cite {Anselmi}. However, it is also not causal 
since the quadratic curvature invariant  terms --  when stand alone -- 
generate the Ostrogradsky ghosts,  whose removal leads to acausalities.  
In  contrast with this, our approach assumes that gravity is to be completed at some high scale by new physics 
(such as string theory). Thus there  should be  an infinite number of curvature terms  
in the classical  action in addition to the Einstein-Hilbert term. 
These terms  do not result from loops, but from a tree-level expansion 
(say, the $\alpha'$ expansion in  string theory). Moreover,  we  argue that there are no 
low energy loop corrections in the gravity sector since $\hbar q \to 0$.  Thus, the SM loops will just renormalize 
already pre-existing $R^2$  and higher terms, but the series in curvature invariants should not be truncated 
at any finite order when it comes to the discussion of the Ostrogradsky ghost; 
hence, neither the existence of such ghosts,  nor the  acausality  can be  established  in our case.}

\subsection{Gravity Loops} 

A quantum theory of gravity  is likely  to come with new degrees of 
freedom  at the energy scale $M_{\rm QG}$, as does string theory.  
However,  even before the full  quantum theory of gravity is explored, there is a more  
immediate  issue  within the  low energy  effective theory:  
the quantum gravity corrections should  be  small  at momentum/energy  scales  below  $M_{\rm QG}$ 
for a classical treatment to  be  sensible.   For instance, if  Einstein's  gravity is regarded as 
part of a low energy quantum effective field theory,   the loops only generate  higher dimensional 
operators that make small contributions at momenta/energies below $M_{\rm QG}$.  Hence,
all the classical predictions of the Einstein  theory at length/time scales well above 
$M_{\rm QG}^{-1}$ are intact, to a good approximation.  We should then strive to make 
sure that the quantum corrections are small at large length/time scales in the proposed framework. 
Given that the classical  action  has an unusual form in the present case, 
one first needs to set  the rules   of calculation for the gravity loops, no matter how small their value 
could be.  These rules were outlined in \cite {Gabadadze:2014rwa}; we 
discuss  them in more detail here, especially in the context of Ref. \cite {Anselmi}.

Following \cite {Gabadadze:2014rwa}, we consider the path integral for gravity,
\be
Z_g= {\rm const} \times \int  d\mu(g) d \mu(F) d\lambda\, dq\,  {\rm exp} \left \{{i \over \hbar} \left ( 
\frac{1}{q}(S_{\rm mGR}+\Gamma_{\rm eff})+ \lambda((V_g/V_F)-q) + S_F \right ) \right \}\,,
\label{Zg}
\ee
where $S_{\rm mGR}$ is the massive gravity action for the $g$-universe,  the two measures 
$ d \mu(g) $ and  $ d \mu(F)$  include the gauge fixing conditions and 
the Faddeev-Popov determinants for both $g$ and $F$, and $\Gamma_{\rm eff}$ is the real part 
of  the 1PI effective action  for the SM fields discussed in the previous subsection. 
$\Gamma_{\rm eff}$ contains  real parts of all possible Green's functions of the SM fields, 
with gravity treated as an external field.  The imaginary parts for these  Green's functions can be 
restored from the real parts of the  lower order Green's functions, by using the optical theorem 
\cite {Anselmi}.  Thus, $\Gamma_{\rm eff}$ contains in principle all the information  about the 
quantized SM  in classical gravitational field. 

The above path integral defines an algorithm, albeit unconventional  and a bit cumbersome,
for calculating quantum loops:  to reiterate, the SM loops are done in a conventional way using $\hbar$ 
while treating  $g$ and $F$ as  external classical fields;  this gives rise  to  $\Gamma_{\rm eff}$ defined in 
the previous subsection. Then, the gravity loops are done using (\ref {Zg}); these loops would contain 
only real parts of the SM Green's functions since $\Gamma_{\rm eff} = {\rm Re}~\!\Gamma_{\rm {1PI}}$. 
The respective imaginary parts  can be obtained from the optical theorem  for the SM loops;  in our 
case they should  be added to the real parts of the SM Green's functions that 
would appear in the order-by-order expansion  in (\ref {Zg}) (one  can of course get  both the 
real and imaginary parts  of those Green's functions if one calculates $\Gamma_{\rm {1PI}}$; it's also 
well know that the solutions obtained from $\Gamma_{\rm {1PI}}$ are gauge independent).

In this scheme, the   parameter  $\hbar q$ may be regarded as a second  Planck constant that 
governs the gravity loops at low energies. Since the classical solutions in our case are such 
that $\hbar q \to 0$,  and hence no gravity loop contributions survive,  it is then 
self-consistent to use classical gravity equations (\ref {GR+SM}).\footnote{
The obtained action, $S_{\rm gravity}+ \Gamma_{\rm eff}$, can be used to develop 
a tree-level nonlinear perturbation theory to account for all the known classical 
gravity, astrophysics, and cosmology effects. One can  try to go further and address a more ambitious 
program of defining the theory for $q\neq 0,q\neq 1$. This is not our goal here, but we make a few comments on such
a prospect: In that case,  
the mixed gravity-SM quantum loops  can be done  by integrating  in (\ref {Zg}) further 
with respect to  the SM ``classical  fields" upon which $\Gamma_{\rm eff}$ depends.  This integration   
leads to a ``double"  quantization of the SM fields  -- the primary quantization  of the SM fields 
being done with $\hbar$, and  the  secondary  quantization  of the ``classical" 
SM fields appearing in the effective action,  being done with the rescaled 
constant $\hbar q$.  As long as  $\hbar q \to 0$,
the secondary  quantization does not modify the results of the primary 
quantization of the SM fields.  However, the procedure may lead to 
unconventional results if  $q\neq 0, q\neq 1$, and this needs to be studied  further. 
All we can say here is that in the limit $\hbar q \to 0$
the optical theorem is valid  even for gravity and mixed loops 
order by order in $\hbar q$, given that one incorporates the optical theorem 
for the SM Green's functions to restore their respective  imaginary parts as discussed 
above;  the optical theorem for the gravity and mixed loops  would relate   tree  
amplitudes to the respective loop amplitudes, even though the loop-generated counterterms  
to the effective action tend to  zero as $\hbar q\to 0$. For instance, consider a one-loop  
diagram that describes a  self-energy correction due to a graviton 
to one of the  ``classical" SM fields appearing in $\Gamma_{\rm eff}$. This diagram has  two vertices, 
each  proportional to  $1/\hbar q$,  and two propagators, each proportional to $\hbar q$, hence the diagram is ${\cal O}(1)$,  if  we choose to normalize the one-particle states so that the 
external lines in the diagram carry no powers of $\hbar q$.   
The imaginary part of this diagram, which is also ${\cal O}(1)$, 
describes a classical process of emission of a graviton  by a SM particle. This amplitude 
is certainly nonzero in the $\hbar q \to 0$ limit. However, the  loop diagram itself 
does not introduce any counterterms in the   partition function --  
being ${\cal O}(1)$, it  is suppressed by $\hbar q$, as compared to the leading tree-level 
terms in the path integral that are  of the order $1/\hbar q$. We will not pursue here 
these arguments further, and will restrict to the limit $\hbar q \to 0$   for the reminder of the 
paper.  We  thank V.P. Nair for useful discussions  of the above issues.}

With the quantization process defined as above, one needs to worry 
about two types of quantum corrections that may potentially ruin the 
classical solution of the cosmological constant problem in  the $g$-universe:
(A)  The action  (\ref {ac1}) and (\ref {acmain})  appear to suggest that there 
should be novel nonlocal interactions in the $g$-universe, arising due to the 
product of two integrals in  $V_g^{-1}S_{\rm mGR}$; 
 (B) The form of the effective action in   (\ref  {expA}) and ({\ref {Zg})   
 raises the  question of whether new polynomial terms   
 of $\lambda$ and $q$  may be generated by quantum corrections, 
 and whether these new terms can spoil the classical solutions. 
 
To address the point (A), let us decompose the metric   
as a background, $g_b$,  and its fluctuation, schematically $g =g_b +h$. The 
inverse volume factor, $V_g^{-1}$, multiplying the  action $S_{\rm mGR}$ , 
can then  be expanded  as follows: $V_g^{-1} = V_b^{-1} - V_b^{-2} H_h+...$, where $V_b = \int d^4 x\sqrt{|g_b|}$  and 
$H_h = \int d^4 x\sqrt{|g_b|} h/2$.  The  term,   $- V_b^{-2} H_hS_{\rm mGR}$, as well as  the other 
terms containing higher powers of $h$, will produce new unconventional interaction vertices 
at both the tree level and  in the loops.  When sandwiched between various states, 
the tree-level terms either give trivially zero, or are suppressed by extra powers of the inverse volume, 
$V_b^{-1}$ as long as all the fields involved decay at spatial/time infinity;  
they can also give rise to amplitudes  that do   not 
correspond to any scattering process.\footnote{Similar amplitudes are also present in ordinary theories at higher orders 
in $1/\hbar$. For instance,  in a theory of a massive scalar $\phi$, the tree amplitude obtained by 
sandwiching the operator $\hbar^{-2} \int d^4 x \,m^2 \phi^2(x) \times \int d^4y\, m^2 \phi^2 (y)$ between 
two two-particle states does not correspond to any scattering process.}  As to the loops arising from these  novel vertices, they 
will be suppressed by powers of  $V_b^{-1}$, in addition to being governed by the vanishing 
effective Planck constant,  $\hbar q$.  

To address the point (B),  let us rewrite the partition function 
in  a slightly different  form:
\be
\int \, d \mu(g) d\mu(F)\, d\lambda \, d p \, {\rm exp}  \left \{{i \over \hbar} \left ( 
\frac{V_F}{p}(S_{\rm mGR}+\Gamma_{\rm eff})+ \lambda(V_g -p) + S_F\right ) \right \}\,, 
\label{p}
\ee
where  we  integrate  with respect to $p$  and $\lambda$. Since  there is no Wick contraction 
between $g$ and $F$, the multiplier  $V_F$ in front of $S_{\rm mGR}+\Gamma_{\rm eff}$
does not produce any new vertices in the $g$-universe. Hence, this multiplier can be regarded as a pure 
number from the point of view of the $g$-universe.   The new nonlocal vertices described in question (A)  
are now encoded in the conventional-looking term $\lambda V_g$.  These  new vertices 
will modify  the calculations, but   the 
modifications  are  straightforward to take into account. To do so, let us further rewrite the partition 
function,
\be
\int \, d \mu(g) d\mu(F)\, d\lambda \, d p \, {\rm exp}  \left \{
\frac{i}{{\tilde \hbar}} \left ( S_{\rm mGR}+ \Gamma_{\rm eff} + {\tilde \lambda}(V_g -p) \right) + 
{i \over \hbar} S_F \right \}\,, 
\label{p1}
\ee
where ${\tilde \hbar} =\hbar p/V_F$,  and  ${\tilde \lambda} = \lambda p/V_F$.
Then,  we consider loop corrections about a flat $g$-background, which is a solution 
for arbitrary CC in the $g$-universe.  Due to these corrections   there will be additional terms proportional 
to positive powers of $\tilde \hbar$, which need to be included in the effective action  
for gravity. The  most dangerous  of these terms would be proportional to 
\be
{\tilde \hbar}^{k+1} {\tilde \lambda}^{l+2}\,, \quad {k,l =0,1,2,3, ...}~\,
\label{indterms}
\ee
However, it is straightforward  to see that on the classical solutions of the original theory, 
all  these terms vanish when  $V_g/V_F\to 0$. 

We end this section by  commenting on another type of corrections that are  likely to 
exist in the theory.  The action $S_{\rm mGR}$ contains the Einstein-Hilbert term, but it may also contain 
higher derivative terms, such as  higher powers of curvature invariants,
that  are not necessarily induced by quantum loops. These terms can arise in 
a putative UV completion of  gravity.  In string theory, for instance, these terms 
would be due to the $\alpha^\prime$ corrections. Such terms  will not spoil 
any of our arguments as long as we are considering  energies and momenta 
below the scale by which these higher dimensional operators are  suppressed 
in comparison  with the EH term. Since the latter scale 
should be expected to be of order $M_{\rm QG}$, the  effects  
of these higher dimensional operators are then negligible  for the SM fields at and below 
$M_{\rm SM}$.  On the other hand, the higher dimensional operators will  be relevant 
for physics at very short distances. One can hope that these 
operators will smooth out the short length/time singularities  of 
certain classical solutions,  e.g.,  black holes, or cosmological solutions. Similar considerations apply to  
higher dimensional operators  that are also expected to appear in $S_F$.

\section{On Strong Coupling in Massive Gravity}\label{secstrong}

The diffeomorphism invariant action for massive gravity was built in 
\cite{deRham:2010ik,deRham:2010kj}, and  was presented  
in (\ref {mg1}).  The specific structure of the potential guarantees that the theory propagates 
only 5 degrees of freedom \cite{deRham:2010ik,deRham:2010kj,Hassan:2011hr,DeffayetEsposito,
Mirbabayi:2011aa,Kugo,Gao}.  
It is not guaranteed,  however,  that the above structure is preserved by loop 
corrections. We note  that the coefficients 
$\alpha_{3,4}$ in (\ref {mg1}) get renormalized only multiplicatively \cite {dRGHP}, i.e.  if set to zero they 
remain zero, but the loops would  in general induce other terms such as 
$\det_3(\mathcal{K})^2$ or  $\det_3(\mathcal{K}) \times \det_4(\mathcal{K})$, etc., that 
would reintroduce the sixth ghostly degree of freedom at a certain energy 
scale.

To see this  more explicitly, we take the so-called decoupling limit
$\mpl \to \infty$, $m\to 0$, with $\Lambda_3 = (\mpl m^2)^{1/3}$ fixed \cite{AGS}. In this limit, the  five 
polarizations of the massive graviton  acquire  their individual  identities as the helicity 
$\pm 2$, helicity $\pm 1$ and helicity 0 states. For simplicity, we focus on the helicity-0 
state, which we denote by $\pi$.  The massive gravity  action  in the decoupling limit
contains the following terms for the $\pi$ field:
 \be
S_\pi = \int d^4 x  \left ( -{1\over 2} (\d \pi)^2 + \alpha {(\d \pi)^2 \square \pi \over \Lambda_3^3} + 
{2 \alpha^2\over 3}{(\d \pi)^2  
((\d \d \pi)^2-(\square \pi)^2 ) \over \Lambda_3^6} +... \right )\,.
\label{GalS}
\ee
These are the so-called Galileon terms \cite {Nicolis}; they are special since they 
don't generate higher than two derivatives in the equation of motion, thus 
retaining only one degree of freedom in the $\pi$ theory 
on an arbitrary background.  

On the other hand, it is evident from the above action that interactions of the $\pi$  field 
are described by irrelevant operators, that  become strong  at 
the scale $\Lambda_3$. If this theory  is quantized in a conventional way, the renormalized Lagrangian 
would include an infinite number of counterterms with higher derivatives, such as higher powers 
of $\partial\partial \pi$. 

These new terms, if present, would introduce a ghost at the scale $\Lambda_3$.
One  way to deal with this  problem  is  to regard the action (\ref {GalS})
as an  effective action valid below the scale $\Lambda_3$, above which it needs 
to be completed by some unknown new physics.  The effects of the new physics in the low energy theory
would manifest themselves as an infinite series
of higher derivative operators  suppressed by  $\Lambda_3$. 
The appearance of the ghost at that scale  could then  be attributed  to  the artificial 
truncation of the series at  a finite order.  Such an approach is 
possible, and has so far been often adopted. It  calls for an answer to the  question 
as to what is the completion of massive gravity at the 
scale $\Lambda_3$,  especially given that the latter  is so 
much smaller than $M_{\rm Pl}$ (see \cite{Kurt,Claudia},
and references therein).

In our approach, however,  such a question does not arise. This is because  the massive 
gravity Lagrangian is not quantized  with $\hbar$,  but instead with $\hbar q$.  The appropriate part of the 
path integral reads schematically as
\be \label{GalZ}
{\rm exp} \left (  {i\over \hbar q}  S_\pi \right )\,.
\ee
Thus, every $\pi$  propagator will be proportional to $\hbar q$ and every 
$\pi$ vertex, be it cubic or quartic, to  $1/ \hbar q$.  If we normalize one-particle states 
in a way that the standard loop expansion is an expansion in powers of the respective 
Planck constant, then all the loop-generated counterterms  will be proportional to 
positive powers of $\hbar q$, and would vanish in the setup considered here.

We note that this does not obviate the strong coupling scale $\Lambda_3$, 
but only renders the full quantum theory with  a finite number of tree-level 
terms, so the theory is still strongly coupled.   While there are no quantum mechanical constraints 
imposed on the amplitudes,  since $\hbar q \to 0$, a resummation of the  classical nonlinear
diagrams would be needed to account for the Vainshtein effect \cite {Arkady}
(see also \cite {DDGV}).  It is also  relevant to note that 
the   tree amplitudes define  loop level amplitudes via the optical theorem which remains valid  order-by-order,  
but  the effective action receives no loop-generated counterterms.  This statement is trivial 
if we restrict ourselves to the $\pi$-sector only -- that sector has only one effective Planck's constant, $\hbar q$,  
hence the reason for validity of the optical theorem is identical to that in a conventional theory.\footnote{
However,  the situation changes when one couples the $\pi$-sector with the SM fields that are now 
quantized with a different Planck's constant, $\hbar q$.  The validity of the optical theorem for $q \neq 0, q\neq 1$ 
is plausible,  but has not been demonstrated,  as discussed in Footnote 14.}
To see how this works in more detail, consider the  $2\to 2$ scattering amplitude at the one-loop level 
that is due to the Wick contraction of two quartic  Galileon terms.   We recall that every  propagator will
 be proportional to $\hbar q$,  and every tree level vertex  will be proportional to   
 $1/ \hbar q$.   Moreover, if we normalize one-particle states 
in a way that  the external lines in Feynman diagrams carry no factors of $\hbar q$,  
then, the one loop diagram for  the  $2\to 2$ scattering amplitude  that comes from a 
contraction of two quartic Galileons would be ${\cal O}(1)$. The imaginary part of this diagram is 
also of order ${\cal O}(1)$, and this is equated  to a square of the tree-level diagram, times
two powers of $\hbar q$ coming from the insertion of the intermediate states, ensuring
the validity of the optical theorem.    While the imaginary part of 
this one-loop diagram would satisfy the optical theorem, the diagram itself  would not give rise 
to a counterterm in the  exponent of the path integral,  since this  counterterm   would  be   
of order ${\cal O}(1)$, while the  existing tree-level terms in the exponent  are of order $1/\hbar q$.  As to the 
$2\to 2$  amplitude itself, it also receives a  dominant tree-level contribution from the  contraction of two 
qubic Galileon terms, and the corresponding amplitude -- with the normalization chosen here  for  
the external state --  is  of order,  $\hbar q \times (1/\hbar q)^2 =1/\hbar q$.

Thus,  the full quantum effective action for the $\pi$ field has  as  few terms as the 
classical action. All the calculations done in this theory --  although nonperturbative -- 
are  exact.  In other words,  all the Feynman diagrams are defined  just by the tree-level  
action. For  external classical sources, the Vainshtein mechanism  \cite {Arkady}
will postpone the strongly coupled regime  to  energy/momentum scales higher 
than $\Lambda_3$ (see \cite {Babichev} for a review) .

\section{Comments and Outlook}\label{seclast}

This section consists of a  few  comments on topics that are 
somewhat disconnected  from each other,  but  might be  
helpful  to  further explore the present proposal.

First we note that  since $\hbar q\to 0$, low energy gravity is essentially classical, 
and therefore, there will  be no tensor modes generated by quantum fluctuations during inflation.
The scalar mode will still be generated  since the inflaton  is quantized in a conventional way,
and even though the scalar fluctuation is a mixture of inflaton and  
metric fluctuations,  one could choose  a gauge in which  the scalar perturbation 
is due entirely to an inflaton\footnote{One could see this also  without choosing a gauge 
but by diagonalizing and rescaling the system of inflaton and metric perturbations 
in the scalar sector.}; thus, our framework would not change the inflationary 
predictions  for  the scalar perturbations.  However,  if the primordial quantum tensor 
modes are discovered this would rule out our proposal  with  $\hbar q\to 0$.

In the context of the present work, we argued that it could be 
advantageous to consider massive gravity as an agent driving 
the accelerated expansion of the universe.  
However, as we have mentioned, massive gravity is not the only way. One could easily imagine a 
quintessence field protected by symmetries, 
as in the  pseudo-Nambu-Goldstone boson (PNGB) scenario of \cite {JoshF}, 
to give rise to dark energy.   It is  straightforward to see  that all our requirements would remain intact
if we assumed that  the PNGB gave rise to dark energy in the $g$-universe on the boundary 
of AdS$_5$.  While  the  $g$-universe in  that  case   would tend  to 4D Minkowski space in the infinite future,
the ratio $V_g/V_F$ would still tend to zero since  $V_F$ (= volume of AdS$_5$) is infinitely
larger than the volume of infinite 4D Minkowski space. While the PNGB model 
would give a redshift dependent equation of state for dark energy, massive gravity and its extensions
give the redshift-independent equation,  $p =-\rho$.  

It is worth  noting that the action  (\ref{modaction}) allows for straightforward Euclidean extensions, 
since the  standard rotation to Euclidean times, $t_{\rm E}=it$,  done in both universes simultaneously,  
amounts to the usual Euclidean transition in the generating functional.  Thus, the solutions that 
have real Euclidean counterparts (say, maximally symmetric spaces)  can 
be discussed  using the Euclidean path integral.  One could  study  in more detail  the AdS/CFT  
algorithm in Euclidean space, and see how it is affected  when the $V_F/V_g$ factor is 
introduced.\footnote{Without this factor,  the only novelty in our case is that  the full 
boundary action  would become $S_{\rm mGR}[g,\gamma]+S_{\rm{CFT}}$.}

We'd like to comment on  the RS2 model \cite{Randall:1999vf}, where induced 4D gravity exists on the RS brane (set $z=\epsilon>0$  and impose $Z_2$ across the hypersurface 
in Sections \ref{secm1} and \ref{secm2}). This scenario  would allow the boundary metric $\gamma_{ab}$ to be truly dynamical. For instance, its linearized Einstein kinetic term is given in (\ref{kin1}), from which it is clear that a mass scale proportional to $1/\epsilon$ acts as the effective Planck mass for $\gamma_{ab}$. Then, the braneworld will contain a ghost-free bi-gravity theory in the framework of Section \ref{secm1}  \cite{exten2}, or a globally connected bi-metric theory in that of Section \ref{secm2}. However, the value of the CC would need to be fine-tuned in order to fulfill a set of junction conditions that maintain the consistency of the RS2 picture, in which case our solution to the big CC problem would be lost. In retrospect, the fine-tuning that precludes the proposed solution  of the CC problem 
to be applicable in the RS2 scenario,  is not required on the conformal boundary of AdS$_5$  (i.e., when 
$\epsilon \to 0$) because there is no $F$-metric dynamics in the $\epsilon\rightarrow 0$ limit.

Last but not least, it would be interesting to  look into the   vierbein formulation of massive gravity 
\cite {RachelKurt,Potential}  in this scenario:   besides the St\"uckelberg fields  $\varphi^a$,  the vierbein 
formulation  requires a two-index field $\lambda^{a}_{\bar a}$ \cite{Potential}. It would be   fascinating   
if one were able to understand the origin of this field from a broader context.

\section*{Acknowledgements}

We are grateful to A. Kobakhidze and V.P. Nair for useful discussions.    
Both GG  and SY are    supported by  NASA  grant  NNX12AF86G  S06,  and GG also by 
NSF  grant PHY-1316452. GG acknowledges a membership at 
the NYU-ECNU Joint Physics Research Institute in Shanghai.

\appendix
\section{Boundary Fields in AdS/CFT}\label{append1}
In this section, we switch to the Euclidean signature, in tandem with the standard language of AdS/CFT. We present how to find the dual boundary CFT action for bulk graviton. As a precursor, we discuss the massless scalar case first, which is simpler but very similar to the graviton case.
\subsection{Massless Scalar}
The Euclidean AdS$_5$ metric is
\begin{equation}\label{metric1}
ds^2=F_{AB}dx^Adx^B=\frac{l^2}{z^2}(dz^2+d\vec{x}^2).
\end{equation}
We set $l=1$ in the following discussion and only restore it when necessary. The scalar action is
\begin{equation}
S=\frac{1}{2}\int d^4xdz~\sqrt{F}F^{AB}\partial_A\Phi\partial_B\Phi,
\end{equation}
whose variation gives the equation of motion
\begin{equation}
\left(\partial_z^2-\frac{3}{z}\partial_z+\square\right)\Phi(z,x)=0.
\end{equation}
The general solution that decays at the horizon $z=\infty$ is $\Phi\sim z^2K_2(z\sqrt{-\square})\phi_0(x)$. Integrating by parts and using equation of motion,
\begin{equation}\label{A4}
S=\frac{1}{2}\int d^4x\sqrt{\tilde{\gamma}}\Phi\hat{n}\cdot\nabla\Phi,
\end{equation}
where $\tilde{\gamma_{ij}}$ is the induced metric on the boundary and $\hat{n}\cdot\nabla=-z\frac{\partial}{\partial z}.$ To obtain the effective boundary action, we set $z=\epsilon$ and take $\epsilon\rightarrow 0$ in the end. We call the boundary condition $\Phi(\epsilon,x)=\phi_0(x)$, so that
\begin{equation}
\Phi(z,p)=\frac{z^2K_2(pz)}{\epsilon^2 K_2(p\epsilon)}\phi_0(p)
\end{equation}
in the Fourier space.

Plugging this into (\ref{A4}),
\begin{equation}
S=-\frac{1}{2}\epsilon^{-4}\int d^4xd^4y~\phi_0(x)\phi_0(y)\int \frac{d^4p}{(2\pi)^4}e^{ip(x-y)}u\frac{\partial}{\partial u}\ln(u^2K_2(u)),
\end{equation}
where $u=p\epsilon.$ We then expand the integrand in series,
\begin{equation}
u\frac{\partial}{\partial u}\ln(u^2K_2(u)) =-\frac{u^2}{2}+\frac{u^4}{4}\left( \ln 2 -\gamma -\ln u \right)+\cdots .
\end{equation}
Here, $\gamma=0.5772...$ is the Euler-Mascheroni constant. The higher order terms in the series can be ignored because they contain more than four powers of $\epsilon$ and drop out when $\epsilon\rightarrow 0$.
 
After performing the Fourier transform, we have\footnote{The relevant formulae are given in \cite{Hinterbichler:2015pta, Freedman:1991tk}, where the nonlocal expressions are ``differentially regularized" to give well-defined Fourier transform.}
\begin{align}\label{ea1}
S[\phi_0]=&-\frac{\epsilon^{-2}}{4}\int d^4x~\phi_0(x)\square\phi_0(x) \nonumber \\
&+\frac{1}{32\pi^2}\int d^4xd^4y~\phi_0(x)\phi_0(y)\square_x^3\left[ \frac{\ln(|x-y|/\epsilon)}{|x-y|}\right].
\end{align}
The content of the second line can be clarified by introducing an arbitrary and finite mass scale $\mu$,
\begin{align}
\square_x^3\left[ \frac{\ln (|x-y|/\epsilon)}{|x-y|^2}\right]&=\frac{1}{2}\square^3_x\left[ \frac{\ln(|x-y|^2\mu^2)}{|x-y|^2}\right]-\square^3_x\left[ \frac{\ln(\mu \epsilon)}{|x-y|^2}\right] \nonumber \\
&=\frac{1}{2}\square^3_x\left[ \frac{\ln(|x-y|^2\mu^2)}{|x-y|^2}\right]+4\pi^2\ln(\mu \epsilon)\square^2_x\delta(x-y).
\end{align}
The first term is the ``differentially regularized" two-point correlator for the boundary scalar CFT operator, up to a constant multiple:\footnote{See Appendix B of \cite{Hinterbichler:2015pta}. Our formulae are obtained by setting $k=2$ there.}
\begin{align}\label{plugin3}
\frac{\pi^2}{24} \Expect{\O(x)\O(y)}_{\text{reg}}&=-\frac{1}{2^7\cdot 6} \square^3_x\left[ \frac{\ln (|x-y|^2\mu^2)}{|x-y|^2} \right] \\
&= \lim_{\kappa\rightarrow 0} \left\{ \frac{1}{|x-y|^{8+2\kappa}}+\frac{\mu^{2\kappa}}{\kappa}\frac{\pi^2}{2^5 \cdot 6} \square^2_x\delta (x-y) \right\},
\end{align} 
so the well-known $\sim |x-y|^{-8}$ two-point function is recovered \cite{2pointfunc}. The total boundary action is
\begin{equation}
S[\phi_0]=-\frac{1}{2}\int d^4x~\phi_0(x)\left(\frac{\epsilon^{-2}}{2}\square-\frac{\ln (\mu\epsilon)}{4}\square^2 \right)\phi_0(x) -\frac{1}{2}\int d^4xd^4y~\phi_0(x)\phi_0(y) \Expect{\O(x)\O(y)}_{\text{reg}}.
\end{equation}

In the AdS/CFT context, the local terms that diverge in the $\epsilon\rightarrow 0$ limit are unphysical and should be removed through holographic renormalization \cite{hologrenorm}. Since the regularized correlator only differs from the unregularized one by local terms, specific renormalization schemes can be constructed to yield either form of the correlator. Still, we note that the differentially regularized expression of the two-point function emerges automatically in the calculation.

\subsection{Graviton}
Now we can look at the boundary theory induced by a free graviton from the AdS$_5$ bulk. The complete action is \cite{Liu:1998bu,Mueck:1998ug}
\begin{equation}
S=-\int_{M} d^4xdz~\sqrt{F}(R+12)-2\int_{\partial M}d^4x~\sqrt{\tilde{\gamma}}K+6\int_{\partial M}d^4x~\sqrt{\tilde{\gamma}},
\end{equation}
up to a functional of $\tilde{\gamma_{ij}}.$ We set the Planck scale $M_5$ to unity for convenience, and use $i,j,...=1,2,3,4$ to denote the coordinates transverse to $z$, which are raised by $\tilde{\gamma^{ij}}$ on the boundary.

The graviton is defined perturbatively by $F_{AB}= F_{AB}^{(0)}+h_{AB}$ with background solution $ F_{AB}^{(0)}$. We work in the gauge $h_{0B}=0.$ In the equations of motion, we use $h^i_j=\tilde{\gamma}^{ik}h_{kj}$ as the dynamical field, since it is related to the boundary condition in a simple way, $h^i_j|_{z=\epsilon}\equiv \chi^i_j=h^0_{ab}$, where $h^0_{ab}$ is the Dirichlet boundary data.\footnote{Here, $a,b,... =1,2,3,4$ are raised with $\delta^{ab}$. In the case that the boundary condition is related to the fiducial metric of massive gravity, the fiducial metric is $f_{\mu \nu}=(\eta_{ab}+h^0_{ab})\partial_{\mu}\varphi^a\partial_{\nu}\varphi^b$.} This allows the latin indices to be raised and contracted with $\delta^{ij}$ on the boundary, which we will do from now on. The equations of motion are \cite{Mueck:1998ug,Arutyunov:1998ve}
\begin{equation}\label{eom2}
\partial_z^2h_j^i+\square h^i_j-\frac{3}{z}\partial_zh^i_j-\frac{1}{z}\partial_z h\delta^i_j+\partial^i\partial_jh-\partial^i\partial_lh^l_j-\partial_j\partial^lh^i_l=0,
\end{equation}
with constraints
\begin{equation}
\square h-\partial_i\partial^j h^i_j-\frac{3}{z}\partial_zh=0,
\end{equation}
\begin{equation}
\partial_z(\partial_ih-\partial_jh^j_i)=0.
\end{equation}

(\ref{eom2}) implies that the transverse-traceless component of $h^i_j$ satisfies the free massless scalar equation of motion. We can then write the solution of $h^i_j$ in Fourier components \cite{Mueck:1998ug}, 
\begin{align}\label{h1}
h^i_j(z,x)=&\int\frac{d^4p}{(2\pi)^4}e^{ipx}\left[ \bar{\hat{h}}^i_j\frac{z^2K_2(pz)}{\epsilon^2K_2(p\epsilon)}-\frac{z^2-\epsilon^2}{6}\left(p^ip_j\hat{h}-\frac{p^ip_jp^kp_l}{p^2}\hat{h}^l_k \right) \right. \nonumber \\
&\left. +\frac{p_jp^l}{p^2}\hat{h}^i_l+\frac{p^ip_l}{p^2}\hat{h}^l_j-\frac{p^ip_jp^kp_l}{p^4}\hat{h}^l_k+\frac{1}{3}\left(\delta^i_j-\frac{p^ip_j}{p^2}\right)\left( \hat{h}-\frac{p^kp_l}{p^2}\hat{h}^l_k \right)  \right],
\end{align}
where $\hat{h}^i_j(p)$ is the Fourier transform of $h^i_j|_{z=\epsilon}=\chi^i_j,$ and
\begin{equation}\label{h2}
\bar{\hat{h}}^i_j=\hat{h}^i_j-\frac{p_jp^l}{p^2}\hat{h}^i_l-\frac{p^ip_l}{p^2}\hat{h}^l_j+\frac{p^ip_jp^kp_l}{p^4}\hat{h}^l_k-\frac{1}{3}\left(\delta^i_j-\frac{p^ip_j}{p^2}\right)\left( \hat{h}-\frac{p^kp_l}{p^2}\hat{h}^l_k \right).
\end{equation}
Using equation of motion and integrating by parts, the action becomes \cite{Liu:1998bu,Mueck:1998ug,Arutyunov:1998ve}
\begin{equation}\label{a2}
S=-\frac{1}{4}\epsilon^{-3}\int d^4x (h^j_i\partial_zh^i_j-h\partial_zh)|_{z=\epsilon}.
\end{equation}
We can substitute (\ref{h1}) and (\ref{h2}) into this expression to obtain an effective action for $\chi^i_j$. The nonlocal terms from this action are worked out in \cite{Liu:1998bu,Mueck:1998ug}. Here, we give the local terms and the differentially regularized two-point function. The total effective action is
\begin{equation}
S[\chi]= S_{\partial^2}+S_{\partial^4}+S_{\text{nonlocal}},
\end{equation}
analogous to the massless scalar case. To obtain the results below, 4D integration by parts is used when necessary.

$S_{\partial^2}$ receives contribution from both terms in (\ref{a2}):
\begin{equation}\label{d2}
S_{\partial^2}=\frac{\epsilon^{-2}}{2}\int d^4x~\left[-\frac{1}{4}\chi^j_i\square\chi^i_j+\frac{1}{2}\chi^j_i\partial_j\partial^l\chi^i_l+\frac{1}{4}\chi\square\chi-\frac{1}{2}\chi^j_i\partial^i\partial_j\chi \right],
\end{equation}
which has the structure of the kinetic terms in linearized Einstein gravity, as expected.

Both $S_{\partial^4}$ and $S_{\text{nonlocal}}$ come from the first term in (\ref{a2}) only. Like the massless scalar case, the arbitrary mass scale $\mu$ is introduced to the separate these terms. In particular,
\begin{equation}\label{d4}
S_{\partial^4}=\frac{\ln (\mu \epsilon)}{16}\int d^4x~\chi^j_i D^{ik}_{jl}\chi^l_k,
\end{equation}
where the fully symmetrized differential operator $D^{ik}_{jl}$ is
\begin{align}\label{d41}
D^{ik}_{jl}=&\frac{1}{2}\left(\delta^{i}_l\delta^k_j+\delta^{ik}\delta_{jl}-\frac{2}{3}\delta^i_j\delta^k_l\right)\square^2+\frac{2}{3}\p^i\p_j\p^k\p_l \nonumber \\
&-\frac{1}{2} \left( \delta^i_l\p_j\p^k+\delta^{ik}\p_j\p_l+\delta^k_j\p^i\p_l+\delta_{jl}\p^i\p^k-\frac{2}{3}\delta^k_l\p^i\p_j-\frac{2}{3}\delta^i_j\p^k\p_l \right)   \square. 
\end{align}
Also,
\begin{equation}\label{nl1}
S_{\text{nonlocal}}=\frac{1}{128\pi^2}\int d^4xd^4y~\chi^j_i(x)\chi^l_k(y) \square_xD^{ik}_{jl}(x)\left[ \frac{\ln( |x-y|^2 \mu^2 )}{|x-y|^2}\right].
\end{equation}
We can extract from this the regularized two-point function for the graviton CFT operator, 
\begin{equation}\label{TT}
\Expect{\T^i_j(x)\T^k_l(y)}_{\text{reg}}=-\frac{1}{64\pi^2}\square_xD^{ik}_{jl}(x)\left[ \frac{\ln( |x-y|^2 \mu^2 )}{|x-y|^2}\right].
\end{equation}

We must show that this is consistent with the well-known unregularized expression \cite{Liu:1998bu,Mueck:1998ug},
\begin{align}\label{nl2}
\Expect{\T^i_j(x)\T^k_l(y)}=~&\frac{20}{\pi^2}\frac{1}{|x-y|^8} \nonumber \\
&\times \left[\frac{1}{2}J^i_l(x-y)J^k_j(x-y)+\frac{1}{2} J^{ik}(x-y)J_{jl}(x-y)-\frac{1}{4}\delta^i_j\delta^k_l \right],
\end{align}
where $J^i_j(x)=\delta^i_j-2\frac{x^ix_j}{x^2}.$ It is straightforward to do so. From Appendix B of \cite{Hinterbichler:2015pta} (also Appendix A of \cite{Freedman:1991tk}), we have the identities
\begin{equation}
\lim_{\kappa\rightarrow 0} \left\{ \frac{1}{|x-y|^{4+2\kappa}}+\pi^2\frac{\mu^{2\kappa}}{\kappa}\delta (x-y) \right\}=-\frac{1}{4} \square_x\left[ \frac{\ln (|x-y|^2\mu^2)}{|x-y|^2} \right],
\end{equation}
\begin{equation}
\lim_{\kappa\rightarrow 0} \left\{ \frac{1}{|x-y|^{6+2\kappa}}+\frac{\pi^2}{8}\frac{\mu^{2\kappa}}{\kappa}\square_x\delta (x-y) \right\}=-\frac{1}{32} \square^2_x\left[ \frac{\ln (|x-y|^2\mu^2)}{|x-y|^2} \right].
\end{equation}
Plugging the non-divergent nonlocal parts of these and (\ref{plugin3}) (terms with negative powers of $|x-y|$) into (\ref{TT}) gives (\ref{nl2}) after some work. We note that this is a more rigorous calculation than the result in \cite{Mueck:1998ug}, which arrives at the same expression but does not address the subtlety of regularizing  
Fourier transform near poles.


\begin{thebibliography}{99}

\bibitem{Weinberg:1988cp} 
  S.~Weinberg,
  Rev.\ Mod.\ Phys.\  {\bf 61}, 1 (1989).
  
\bibitem{Polchinski:2006gy} 
  J.~Polchinski,
  hep-th/0603249.

\bibitem{Tseytlin:1990hn} 
  A.~A.~Tseytlin,
  Phys.\ Rev.\ Lett.\  {\bf 66}, 545 (1991).

\bibitem{Linde:1988ws} 
  A.~D.~Linde,
  Phys.\ Lett.\ B {\bf 200}, 272 (1988).

\bibitem{Gabadadze:2014rwa} 
  G.~Gabadadze,
  Phys.\ Lett.\ B {\bf 739}, 263 (2014)
  [arXiv:1406.6701 [hep-th]].

\bibitem{ArkaniHamed:2002fu} 
  N.~Arkani-Hamed, S.~Dimopoulos, G.~Dvali and G.~Gabadadze,
  hep-th/0209227.

\bibitem{Davidson}
A.~Davidson and S.~Rubin,
  Class.\ Quant.\ Grav.\  {\bf 26}, 235019 (2009)
  [arXiv:0905.0661 [gr-qc]].

\bibitem{deRham:2010ik} 
  C.~de Rham and G.~Gabadadze,
  Phys.\ Rev.\ D {\bf 82}, 044020 (2010)
  [arXiv:1007.0443 [hep-th]].
  
\bibitem{deRham:2010kj} 
  C.~de Rham, G.~Gabadadze and A.~J.~Tolley,
  Phys.\ Rev.\ Lett.\  {\bf 106}, 231101 (2011)
  [arXiv:1011.1232 [hep-th]].
  
  \bibitem{Kurt}
  K.~Hinterbichler,
  Rev.\ Mod.\ Phys.\  {\bf 84}, 671 (2012)
  [arXiv:1105.3735 [hep-th]].
  
  \bibitem{Claudia}
  C.~de Rham,
  Living Rev.\ Rel.\  {\bf 17}, 7 (2014)
  [arXiv:1401.4173 [hep-th]].

\bibitem{Fred}
  A.~S.~Goldhaber and M.~M.~Nieto,
  Rev.\ Mod.\ Phys.\  {\bf 82}, 939 (2010)
  [arXiv:0809.1003 [hep-ph]].

\bibitem{dRGHP}
  C.~de Rham, G.~Gabadadze, L.~Heisenberg and D.~Pirtskhalava,
  Phys.\ Rev.\ D {\bf 87}, no. 8, 085017 (2013)
  [arXiv:1212.4128].

\bibitem{GGKurt} G.~Gabadadze, K.~Hinterbichler, D.~Pirtskhalava and Y.~Shang,
  Phys.\ Rev.\ D {\bf 88}, no. 8, 084003 (2013)
  [arXiv:1307.2245].


\bibitem{bigravity1}
N.~Rosen,
  Phys.\ Rev.\  {\bf 57}, 147 (1940).


J. Wess and B. Zumino, in B. Zumino, ``Effective Lagrangians and Broken Symmetries'',  
in Brandeis Univ. Lectures on Elementary Particles
and Quantum Field Theory (MIT Press Cambridge,
Mass.), Vol. 2, 1970, 437. \\


\bibitem{bigravity2}
C.~J.~Isham, A.~Salam and J.~A.~Strathdee,
  Phys.\ Rev.\ D {\bf 3}, 867 (1971).

A.~Salam and J.~A.~Strathdee,
  Phys.\ Rev.\ D {\bf 16}, 2668 (1977).
 


\bibitem{BD}
D.~G.~Boulware and S.~Deser,
  Phys.\ Rev.\ D {\bf 6}, 3368 (1972).


\bibitem{Hassan:2011hr} 
  S.~F.~Hassan and R.~A.~Rosen,
  Phys.\ Rev.\ Lett.\  {\bf 108}, 041101 (2012)
  [arXiv:1106.3344 [hep-th]].


\bibitem{exten2}
 S.~F.~Hassan and R.~A.~Rosen,
  JHEP {\bf 1202}, 126 (2012)
  [arXiv:1109.3515 [hep-th]]. 


\bibitem{KurtRachel} K.~Hinterbichler and R.~A.~Rosen,
  JHEP {\bf 1207}, 047 (2012)
  [arXiv:1203.5783 [hep-th]].


\bibitem{Deffayet} C.~Deffayet, J.~Mourad and G.~Zahariade,
  JCAP {\bf 1301}, 032 (2013)
  [arXiv:1207.6338 [hep-th]].
  
  C.~Deffayet, J.~Mourad and G.~Zahariade,
  JHEP {\bf 1303}, 086 (2013)
  [arXiv:1208.4493 [gr-qc]].
  
  
  

\bibitem{Liu:1998bu} 
  H.~Liu and A.~A.~Tseytlin,
  Nucl.\ Phys.\ B {\bf 533}, 88 (1998)
  [hep-th/9804083].

 \bibitem{exten1}
 S.~F.~Hassan, R.~A.~Rosen and A.~Schmidt-May,
  JHEP {\bf 1202}, 026 (2012)
  [arXiv:1109.3230 [hep-th]].

\bibitem{quasidila}
G.~D'Amico, G.~Gabadadze, L.~Hui and D.~Pirtskhalava,
  Phys.\ Rev.\ D {\bf 87}, 064037 (2013)
  [arXiv:1206.4253 [hep-th]].

\bibitem{Gabadadze:2014kaa}
  G.~Gabadadze, R.~Kimura and D.~Pirtskhalava,
  Phys.\ Rev.\ D {\bf 91}, no. 12, 124024 (2015)
  [arXiv:1412.8751 [hep-th]]. 
  Phys.\ Rev.\ D {\bf 90} (2014) 2,  024029
  [arXiv:1401.5403 [hep-th]]. 
  

  
  
  \bibitem{extensions}
 Q.~G.~Huang, Y.~S.~Piao and S.~Y.~Zhou,
  Phys.\ Rev.\ D {\bf 86}, 124014 (2012)
  [arXiv:1206.5678 [hep-th]];\\
  A.~De Felice and S.~Mukohyama,
  Phys.\ Lett.\ B {\bf 728}, 622 (2014)
  [arXiv:1306.5502 [hep-th]];\\
G.~Gabadadze, K.~Hinterbichler, J.~Khoury, D.~Pirtskhalava and M.~Trodden,
  Phys.\ Rev.\ D {\bf 86}, 124004 (2012)
  [arXiv:1208.5773 [hep-th]];\\
  C.~de Rham, A.~Matas and A.~J.~Tolley,
  Class.\ Quant.\ Grav.\  {\bf 31}, 165004 (2014)
  [arXiv:1311.6485 [hep-th]];\\
   C.~de Rham, A.~Matas and A.~J.~Tolley,
  Class.\ Quant.\ Grav.\  {\bf 32}, no. 21, 215027 (2015)
  [arXiv:1505.00831 [hep-th]];\\
 C.~de Rham and A.~J.~Tolley,
  Phys.\ Rev.\ D {\bf 92}, no. 2, 024024 (2015)
  [arXiv:1505.01450 [hep-th]];\\
 A.~De Felice and S.~Mukohyama,
  Phys.\ Lett.\ B {\bf 752}, 302 (2016)
  [arXiv:1506.01594 [hep-th]].

  
\bibitem{ClassGravity}
 
 T.~W.~B.~Kibble,
  Commun.\ Math.\ Phys.\  {\bf 64}, 73 (1978).

  T.~W.~B.~Kibble and S.~Randjbar-Daemi,
  J.\ Phys.\ A {\bf 13}, 141 (1980).
 
V. P. Frolov and G. A. Vilkovisky, in Quantum Gravity, proceedings
of the Second Moscow Seminar, edited by M.A. Markov
and P. C. West (Plenum, New York, 1983).

A. Vilkovisky, in Quantum Theory of Gravity ,
edited by  S.M. Christensen
(Adam Hilger, Bristol, 1984).
 
\bibitem{InIn} R.~D.~Jordan,
  Phys.\ Rev.\ D {\bf 33}, 444 (1986).
 
\bibitem{Anselmi} 
D.~Anselmi,
  JHEP {\bf 0701}, 062 (2007)
  [hep-th/0605205].


 
 
 
\bibitem{DeffayetEsposito}
  C.~Deffayet, J.~Mourad and G.~Zahariade,
  JHEP {\bf 1303}, 086 (2013)
  [arXiv:1208.4493 [gr-qc]];
  JCAP {\bf 1301}, 032 (2013)
  [arXiv:1207.6338 [hep-th]].


\bibitem{Mirbabayi:2011aa} 
  M.~Mirbabayi,
  Phys.\ Rev.\ D {\bf 86}, 084006 (2012)
  [arXiv:1112.1435 [hep-th]].

\bibitem{Kugo} T.~Kugo and N.~Ohta,
 PTEP {\bf 2014} (2014) 043B04
 [arXiv:1401.3873 [hep-th]]. 
 
 \bibitem{Gao}
 X.~Gao, T.~Kobayashi, M.~Yamaguchi and D.~Yoshida,
  Phys.\ Rev.\ D {\bf 90}, no. 12, 124073 (2014)
  [arXiv:1409.3074 [gr-qc]].

\bibitem{AGS} 
  N.~Arkani-Hamed, H.~Georgi and M.~D.~Schwartz,
  Annals Phys.\  {\bf 305}, 96 (2003)
  [hep-th/0210184].

 \bibitem{Nicolis}
 M.~A.~Luty, M.~Porrati and R.~Rattazzi,
  JHEP {\bf 0309}, 029 (2003)
  [hep-th/0303116];\\
  A.~Nicolis, R.~Rattazzi and E.~Trincherini,
  Phys.\ Rev.\ D {\bf 79}, 064036 (2009)
  [arXiv:0811.2197 [hep-th]].

 \bibitem{Arkady}
  A.~I.~Vainshtein,
  Phys.\ Lett.\ B {\bf 39}, 393 (1972).
  
  \bibitem{DDGV}
  C.~Deffayet, G.~R.~Dvali, G.~Gabadadze and A.~I.~Vainshtein,
  Phys.\ Rev.\ D {\bf 65}, 044026 (2002)
  [hep-th/0106001].

  \bibitem{Babichev}
  E.~Babichev and C.~Deffayet,
  Class.\ Quant.\ Grav.\  {\bf 30}, 184001 (2013)
  [arXiv:1304.7240 [gr-qc]].

   \bibitem{JoshF}
  J.~A.~Frieman, C.~T.~Hill, A.~Stebbins and I.~Waga,
  Phys.\ Rev.\ Lett.\  {\bf 75}, 2077 (1995)
  [astro-ph/9505060].
  
\bibitem{Randall:1999vf} 
  L.~Randall and R.~Sundrum,
  Phys.\ Rev.\ Lett.\  {\bf 83}, 4690 (1999)
  [hep-th/9906064].


   \bibitem{RachelKurt}
  K.~Hinterbichler and R.~A.~Rosen,
  JHEP {\bf 1207}, 047 (2012)
  [arXiv:1203.5783 [hep-th]].
  
  \bibitem{Potential}
  G.~Gabadadze, K.~Hinterbichler, D.~Pirtskhalava and Y.~Shang,
  Phys.\ Rev.\ D {\bf 88}, no. 8, 084003 (2013)
  [arXiv:1307.2245].
      
   
\bibitem{Hinterbichler:2015pta} 
  K.~Hinterbichler, J.~Stokes and M.~Trodden,
  Phys.\ Rev.\ D {\bf 92}, no. 6, 065025 (2015)
  [arXiv:1505.05513 [hep-th]].
  
\bibitem{Freedman:1991tk} 
  D.~Z.~Freedman, K.~Johnson and J.~I.~Latorre,
  Nucl.\ Phys.\ B {\bf 371}, 353 (1992).
      
      \bibitem{2pointfunc}
  E.~Witten,
  Adv.\ Theor.\ Math.\ Phys.\  {\bf 2}, 253 (1998)
  [hep-th/9802150]; \\
  D.~Z.~Freedman, S.~D.~Mathur, A.~Matusis and L.~Rastelli,
  Nucl.\ Phys.\ B {\bf 546}, 96 (1999)
  [hep-th/9804058];\\
  W.~Mueck and K.~S.~Viswanathan,
  Phys.\ Rev.\ D {\bf 58}, 041901 (1998)
  [hep-th/9804035].
  
\bibitem{hologrenorm} 
  S.~de Haro, S.~N.~Solodukhin and K.~Skenderis,
  Commun.\ Math.\ Phys.\  {\bf 217}, 595 (2001)
  [hep-th/0002230];\\
  M.~Bianchi, D.~Z.~Freedman and K.~Skenderis,
  Nucl.\ Phys.\ B {\bf 631}, 159 (2002)
  [hep-th/0112119];\\
  K.~Skenderis,
  Class.\ Quant.\ Grav.\  {\bf 19}, 5849 (2002)
  [hep-th/0209067].
  
\bibitem{Mueck:1998ug} 
  W.~Mueck and K.~S.~Viswanathan,
  hep-th/9810151.
  
\bibitem{Arutyunov:1998ve} 
  G.~E.~Arutyunov and S.~A.~Frolov,
  Nucl.\ Phys.\ B {\bf 544}, 576 (1999)
  [hep-th/9806216].

  
 

 
 
 
  
\end{thebibliography}
\end{document}